\documentclass[12pt]{article}
\usepackage{amssymb,amsmath}
\usepackage{epsfig}
\usepackage{epstopdf}
\usepackage{latexsym}
\usepackage{graphicx}
\usepackage{subfigure}
\usepackage{booktabs}
\usepackage{bbm}

\usepackage[toc]{appendix}

\usepackage{color}
\usepackage{datetime}
\usepackage[
      colorlinks=false,
      linkcolor=darkblue,  
      urlcolor=blue,    
      filecolor=blue,     
      citecolor=red,
linktocpage=true,
      pdfstartview=FitV,
      bookmarksopen=true    
      ]{hyperref}

\ifpdf
\fi

\renewcommand{\theequation}{\arabic{section}.\arabic{equation}}

\def\coeff#1#2{\relax{\textstyle {#1 \over #2}}\displaystyle}
\def\ds{\displaystyle}

\def\cA{{\cal A}}
\def\cB{{\cal B}}
\def\cD{{\cal D}}

\def\cL{{\cal L}}
\def\cM{{\cal M}}
\def\cN{{\cal N}}

\def\cP{{\cal P}}

\def\cR{{\cal R}}

\def\cV{{\cal V}}
\def\cW{{\cal W}}

\def\Neql#1{{\cal N}\!=\!{#1}}
\def\eql{=}
\def\CC{\mathbb{C}}
\def\RR{\mathbb{R}}
\def\PP{\mathbb{P}}

\def\zet{z}
\def\bzet{{\bar z}}

\def\wom{{\omega}}
\def\bzeta{{\bar\zeta}}
\def\ozeta{\overline\zeta {}}
\def\zxi{\xi}

\definecolor{cardinal}{rgb}{0.6,0,0}
\definecolor{darkgreen}{rgb}{0,0.5,0}
\definecolor{golden}{rgb}{0.92, 0.7, 0}
\definecolor{midnight}{rgb}{0, 0, 0.5}
\definecolor{darkblue}{rgb}{0.2, 0, 0.8}

\def\phm{\phantom{$-$}}
\def\bone{{\bf 1}}
\def\bthr{{\bf 3}}

\def\circF{{\buildrel _{\,\,\,\circ} \over F}}
\def\circg{{\buildrel _{\,\,\circ} \over g}}

 \newcommand{\nc}{\newcommand}
\nc{\bea}{\begin{eqnarray}}
\nc{\eea}{\end{eqnarray}}
\nc{\be}{\begin{equation}}
\nc{\ee}{\end{equation}}
 \nc{\tphi}{\tilde{\phi}}
\nc{\non}{\nonumber}

\begin{document}  

\begin{titlepage}
 
\bigskip
\bigskip
\bigskip
\bigskip
\begin{center} 
{\Large \bf  Supergravity Instabilities of Non-Supersymmetric Quantum Critical Points}

\bigskip
\bigskip

{\bf Nikolay Bobev,${}^{(1)}$  Nick Halmagyi,${}^{(2)}$ \\
 Krzysztof Pilch,${}^{(1)}$  and Nicholas P. Warner${}^{(1)}$ \\ }
\bigskip
${}^{(1)}$ Department of Physics and Astronomy \\
University of Southern California \\
Los Angeles, CA 90089, USA  \\
\vskip 5mm
${}^{(2)}$ Institut de Physique Th\' eorique \\
CEA Saclay, CNRS-URA 2306 \\
91191 Gif sur Yvette, France \\
\bigskip
bobev@usc.edu,~nicholas.halmagyi@cea.fr, \\ 
pilch@usc.edu,~warner@usc.edu  \\
\end{center}

\begin{abstract}

\noindent  Motivated by the recent use of certain consistent truncations of M-theory to study condensed matter physics using holographic techniques, we study the $\rm SU(3)$-invariant sector of four-dimensional, $\cN=8$ gauged supergravity and compute the complete scalar spectrum at each of the five non-trivial critical points. We demonstrate that the smaller $\rm SU(4)^-$ sector is equivalent to a consistent truncation studied recently by various authors and find that the critical point in this sector, which has been proposed as the ground state of a holographic superconductor, is unstable due to a family of scalars that violate the Breitenlohner-Freedman bound. We also derive the origin of this instability in eleven dimensions and comment on the generalization to other embeddings of this critical point which involve arbitrary Sasaki-Einstein seven manifolds.  In the spirit of a resurging interest in consistent truncations, we present a formal treatment of the $\rm SU(3)$-invariant sector as a $\rm U(1)\times U(1)$ gauged $\cN=2$ supergravity theory coupled to one hypermultiplet.

\end{abstract}

\end{titlepage}


\tableofcontents

\section{Introduction}
 
Much of the progress in AdS/CMT has been driven by ``phenomenological holography" in which the gravity dual of an interesting condensed matter system is postulated {\it ab initio} (see, for example,  \cite{Gubser:2008px,Hartnoll:2008vx, Hartnoll:2008kx}), without using the more well-established holographic field theories and their gravity duals. In a sense, this sort of approach is reminiscent of Landau-Ginzburg theory but with the important new addition of gravity, whose r\^ole is to provide all the features of holography and, in particular, improve the strong coupling behavior of the effective field theory description.   

It is, of course, likely that many of these more phenomenological models can be realized in either M-theory or IIB supergravity  (see, for example, \cite{Gubser:2008wz, Denef:2009tp, Gubser:2009qm, Gauntlett:2009dn, Gubser:2009gp, Gauntlett:2009bh}), where there is a ``tried-and-true'' holographic dictionary.   It is very important to perform such embeddings not only because of the dictionary but also to make sure that the complete holographic theory does not have other low-mass modes that compromise or destroy the effect that one finds in the reduced or effective field theory.   There are two crucial issues here: consistent truncation and stability.   

The goal is to construct a stable solution of M-theory or IIB supergravity that represents a flow or fixed point of interest.  It is, however, often easier to work in a four- or five-dimensional truncated theory in which many of the complexities of the higher-dimensional supergravity theory are encoded in a scalar potential.   Consistent truncation means that the fields that have been included in the lower-dimensional theory do not source other fields in the higher dimensional supergravity theory and so if one solves a lower-dimensional theory  one is guaranteed to have a solution to the higher-dimensional theory (M-theory or IIB supergravity).  

Stability is rather more subtle.  The supergravity potentials are generally unbounded below and do not even have local minima but merely have critical points at negative values of the potential.  The negative values of the potential lead to anti-de Sitter ($AdS$) vacua that are expected to be dual to non-trivial conformal fixed points in the field theory.  In such a vacuum, one can tolerate a certain amount of negative mass  because the gravitational back-reaction can stabilize the vacuum \cite{Breitenlohner:1982jf}.  To be more precise, suppose that there is a scalar field, $\phi$, in $d$-dimensions with a potential, $\cP(\phi)$,  that has  a critical point at $\phi_0$. Taking the Lagrangian to be
\begin{equation}
e^{-1}{\cal L}\ ~=~ \coeff{1}{2}R - \coeff{1}{2} (\partial_\mu \phi)^2  ~-~ \cP(\phi )\,,
\label{scalLag}
\end{equation}
then the $AdS_d$ vacuum at $\phi_0$ is stable to quadratic fluctuations if  \cite{Breitenlohner:1982jf, Mezincescu:1984ev}
\begin{equation}
{(d-1)(d-2)\over 2}\,\left({\cP''(\phi_0)\over \cP(\phi_0)}\right)~\leq~ {(d-1)^2\over 4}\,.
\label{Bfcrit}
\end{equation}
For more general Lagrangians with more scalars and more complicated kinetic terms, one expands quadratically, normalizes to the form of (\ref{scalLag}) and then applies the Breitenlohner-Freedman (BF) bound, (\ref{Bfcrit}),  to the eigenvalues of the quadratic fluctuation matrix for the potential.  In the holographically dual field theory, violating the bound, 
(\ref{Bfcrit}), shows up as a manifest pathology:  The holographic dictionary shows that such perturbations are dual to operators with complex conformal dimensions.

Supersymmetry guarantees stability through the usual energy bound arguments applied to anti-de Sitter space \cite{Abbott:1981ff, Gibbons:1983aq}.  This implies complete classical and semi-classical stability, and not merely to solutions based upon critical points.  Supersymmetric flow solutions are therefore completely stable.  The problem with imposing supersymmetry is that it also greatly restricts the physics. Indeed, in its simplest form, superfluidity and superconductivity require the formation of a fermion condensate and it is rather unclear whether such a condensation alone could be rendered supersymmetric.  It is possible that there might be some BPS bound that is preserved by the condensate and it would thus be supersymmetric.  There might also be some additional boson condensate, possibly dual to some lattice phonons, so that supersymmetry is preserved.   One might also take the general view, as one frequently does in the study of Yang-Mills theories,  that a supersymmetric condensate, while not describing precisely the physics of interest, could capture some of the universal strong coupling effects of the theory of interest.  Thus supersymmetry is a double-edged sword:  It bestows stability but at the possible cost of losing some of the possible broader physical applications.  It is therefore very valuable to study non-supersymmetric flows and fixed points to determine whether they might be stable and thus describe physically interesting dual field theories.
  
In this paper we will consider the $\rm SU(3)$-invariant sector of gauged $\Neql8$ supergravity in four dimensions and examine the bound, (\ref{Bfcrit}), at all five critical points\footnote{It should be noted that recently other critical points of the $\cN=8$ theory have been found in the impressive work \cite{Fischbacher:2009cj}.}  \cite{Warner:1983vz}  for {\it all} $70$ scalars of the original $\Neql8$  theory. 

Recall the simple argument in \cite{Warner:1983vz} that allows one to study the truncated theory and find extrema of the full scalar potential $\mathcal{P(\phi)}$: consider a subgroup $\rm{G}\subset \rm{SO(8)}$ and denote the  $\rm G$-invariant scalar fields by $\phi_a$ while the others are denoted by $\psi_i$. Now write the full scalar potential of the $\cN=8$ theory as an expansion in the fields $\psi_i$ of the form:
\be
\mathcal{P}(\phi_a,\psi_i)= \sum_{n=0}^{\infty}{\cal P}^{(n)}_{i_1\ldots i_n}(\phi_a)\psi_{i_1}\ldots \psi_{i_n}\,,
\ee
where ${\cal P}^{(n)}_{i_1\ldots i_n}$ is a G-invariant tensor of rank $n$. Similar expansions are performed for all other terms in the supergravity Lagrangian. Since there is no  invariant tensor of rank one, one must have $\mathcal{P}^{(1)}=0$. As a consequence we see that restricting to just the $\phi_a$ fields is indeed a consistent truncation, in particular if $\phi_a=\phi_a^{*}$ is an extremum of $\mathcal{P}^{(0)}(\phi_a)$, then $(\phi_a,\psi_i)=(\phi_a^{*},0)$ is an extremum of $\mathcal{P}(\phi_a,\psi_i)$. However, the stability of this extremum is not guaranteed. To determine stability one must compute $\mathcal{P}^{(2)}(\phi_a)$ (as well as the kinetic terms) and apply the BF bound as outlined earlier. In this work we take  $\rm G=\rm{SU(3)}$ and find that {\it all} non-supersymmetric $\rm SU(3)$-invariant  critical points are unstable within the four-dimensional $\mathcal{N}=8$ theory.  Interestingly enough, these instabilities cannot always be seen within the $\Neql2$ supergravity theory defined by $\rm SU(3)$ invariance:  The instabilities really do involve modes that come from the other $\psi_i$ sector of the complete theory.  
  
In this paper we will pay particular attention to the instability of the non-supersymmetric $\rm SU(4)^-$ critical point, which is one of the examples considered in detail in \cite{Gauntlett:2009dn, Gauntlett:2009bh} in the context of holographic superconductors. This critical point has a well-known uplift to eleven-dimensional supergravity,  the so-called  Pope-Warner (PW) solution \cite{Pope:1984bd}, which is given by an $AdS_4$ compactification with a deformed $S^7$ and internal flux. We study fluctuations around this eleven-dimensional solution, identify the unstable modes and relate them to the unstable scalar modes in four-dimensional $\mathcal{N}=8$ gauged supergravity. We show that the instability in eleven-dimensions is triggered by a flux perturbation represented by a primitive harmonic $(2,2)$ form on the $\mathbb{C}^4$ cone over $S^7$ which transforms in the $\mathbf{20'}$ of $\rm SU(4)$. Through the coupling to the background flux of the PW solution, this form sources other non-trivial flux and metric perturbations. The structure of the unstable modes suggests that the instability may persist also for the more general PW solutions which are obtained by replacing the deformed $S^7$ with more general deformed Sasaki-Einstein manifolds \cite{Pope:1984jj}. 

The $\rm SU(3)$-invariant sector of gauged $\Neql\,8$ supergravity  is an interesting theory in its own right: It is a gauged $\Neql2$ supergravity theory coupled to one vector multiplet and one hypermultiplet.   We will study this model in some detail, and translate between the formulation in terms of the truncated model and the language of special and quaternionic K\"ahler geometry.

In Section \ref{suthreesector} we give complete details of the bosonic action of the truncated $\Neql\,2$ supergravity theory.  In Section \ref{sufoursector} we further specialize this to the $\rm SU(4)^-$ invariant sector and look at the quadratic fluctuations around two of the critical points in this sector. The unstable modes at the $\rm SU(4)^-$ point are studied in detail from the point of view of eleven-dimensional supergravity in Section \ref{sec:instability}. Section \ref{sec:conclude} contains our conclusions. We present details of the stability of all critical points in Appendix \ref{appendixA}.  In Appendix \ref{appendixB} we translate our $\Neql2$ supergravity theory, coupled to a vector multiplet and a hypermultiplet, to the more standard formulation in terms of special and quaternionic K\"ahler geometry. In Appendix \ref{appendixC} we summarize some useful branching rules for irreducible representations of $\rm SO(8)$.

\section{The SU(3)-invariant sector of gauged $\Neql\, 8$ supergravity}
\label{suthreesector}

In this section we compute the complete bosonic action of the SU(3)-invariant sector of gauged $\Neql\,8$ supergravity in four dimensions. While several parts of this action have been computed previously (starting with \cite{Warner:1983vz} where the scalar potential was computed) the full action has never been worked out and we will need it later to precisely compare with results obtained in \cite{Gauntlett:2009bh}.

The eight supersymmetries of the  $\Neql8$ theory decompose as ${\bf 3} \oplus{\bf \bar 3} \oplus {\bf 1} \oplus {\bf 1}$ of $\rm SU(3)$ and so the $\rm SU(3)$-invariant sector has $\Neql2$ supersymmetry. Inside $\rm SO(8)$ the $\rm SU(3)$ commutes with $\rm{U(1) \times U(1)}$ and one of the corresponding  vector fields must be the graviphoton while the other must generate a vector multiplet. As was observed in \cite{Warner:1983vz}, there are six scalars that are $\rm SU(3)$ singlets and two of these must lie in the vector multiplet and so the remainder must lie in a hypermultiplet.  We thus have $\Neql2$ supergravity, coupled to one vector multiplet and one hypermultiplet.  Our purpose in this section is to exhibit, in detail, the embedding of this $\Neql2$ theory in the  $\Neql8$ theory. 

The bosonic sector of the truncated Lagrangian is a sum:
\begin{equation}
\label{lagrangian}
\cL\eql \cL_{Ein.}+\cL_{gauge}+\cL_{kin.}-e\,\cP\,,
\end{equation}
where $\cL_{Ein.}={1\over 2}eR$ is the usual Einstein term,\footnote{We use mostly $+$ convention for the metric signature and denote $e\equiv \sqrt{-g}$.} $\cL_{gauge}$ is the Lagrangian for the gauge fields, which includes couplings to the neutral scalar field, $\cL_{kin.} $ is the kinetic term of the scalar Lagrangian with the minimal coupling to the gauge fields, and $\cP$ is the scalar potential.    If the reader is not interested in the details, $\cL_{gauge}$ is given in (\ref{maxwell})--(\ref{tauans3}),  $\cL_{kin.}$ in  (\ref{sckin})  and $\cP$ in (\ref{poten}).

\subsection{The scalar fields}
\label{scalarsection}

\subsubsection{The $\Neql8$ scalar action}

The scalars of the $\Neql8$ theory lie in the coset $\rm E_{7(7)}/SU(8)$ whose non-compact generators can be represented by a complex, self-dual four-form, considered as a $28 \times 28$ matrix:   $\Sigma_{IJKL}=\Sigma_{[IJ][KL]}$.  That is, one defines a non-compact generator, $G$,  in the 56-dimensional representation of $E_{7(7)}$ (see, Appendix A in \cite{de Wit:1982ig}),
\begin{equation}
\label{esevgen}
G\eql \left(\begin{matrix}
0 & \Sigma_{IJKL} \\[6 pt] \Sigma^{MNKL} & 0
\end{matrix}\right)\,.
\end{equation} 
The components $\Sigma^{IJKL}$ are complex conjugate of $\Sigma_{IJKL}$ and the self-duality constraint is
\begin{equation}
\label{selfdual}
\Sigma_{IJKL}\eql {1\over 24}\epsilon_{IJKLMNPR}\Sigma^{MNPR}\,.
\end{equation}
The exponential map,  $G~\rightarrow~\cV\equiv \exp(G)$, defines coset representatives and determines the scalar vielbein and its inverse,
\begin{equation}
\label{scalviel}
\cV\eql\left( \begin{matrix}
u_{ij}{}^{IJ} & v_{ijKL}\\[6 pt] v^{klIJ} & u^{kl}{}_{KL}\end{matrix}\right)\,,\qquad \cV^{-1} \eql\left( \begin{matrix}
u^{ij}{}_{IJ}  & - v_{klIJ}\\[6 pt] - v^{ijKL} &   u_{kl}{}^{KL} \end{matrix}\right)\,,
\end{equation}
in terms of which the supergravity action is constructed.\footnote{In this subsection capital Latin indices, $I,J,K,
\ldots$, transform under $\rm SO(8)$ and small Latin indices, $i,j,k,\ldots$, transform under $\rm SU(8)$.}  

One then defines a composite $\rm SU(8)$ connection acting on the $\rm SU(8)$ indices according to
\begin{equation}
\label{compconn}
\cD_\mu \varphi^i ~\equiv~ \partial_ \mu\varphi^i  + \coeff{1}{2}  \cB^{\,i}_{\mu \, j} \, \varphi^j\,, 
\end{equation}
and introduces the minimal couplings of the $\rm SO(8)$ gauge fields with coupling constant $g$.  For example
\begin{equation}
\label{covderu}
\cD_\mu u_{ij}{}^{IJ}  \eql \partial _\mu u_{ij}{}^{IJ} -   \coeff{1}{2}  \cB^{\,k}_{\mu \, i} u_{kj}{}^{IJ} - \coeff{1}{2}  \cB^{\,k}_{\mu \, j} u_{ik}{}^{IJ}    - g \left(A_\mu^{KI} u_{ij}{}^{JK} - A_\mu^{KJ}u_{ij}{}^{IK}  \right)\,.
\end{equation}
The composite connections are then defined by requiring that
\begin{equation}
\label{conndefn}
(\cD_\mu \cV)\, \cV^{-1}  ~=~ - \ds\frac{\sqrt{2}}{4}\left( \begin{matrix}
0  & \cA_\mu{}^{ijkl}   \\[6 pt]
  \cA_\mu{}_{\,mnpq} & 0  \end{matrix}\right)   \,.
\end{equation}
More directly, 
\begin{equation}\label{atens}
\cA_\mu{}^{ijkl}\eql -2\sqrt 2\,\left(u^{ij}{}_{IJ}\nabla_\mu v^{klIJ}-v^{ijIJ}\nabla_\mu u^{kl}{}_{IJ}\right)\,,
\end{equation}
where the covariant derivative in \eqref{atens} is only with respect to the $\rm SO(8)$ indices of the scalar vielbeins, that is
\begin{equation}\label{covderv}
\nabla_\mu v^{ijIJ}\eql \partial _\mu v^{ijIJ}-g \left(A_\mu^{KI}v^{ijJK}-A_\mu^{KJ}v^{ijIK}\right)\,,
\end{equation}
and similarly for other fields.

The scalar kinetic term is then given by
\begin{equation}
\label{sckin}
e^{-1}\cL^{\cN\eql 8}_{kin.}\eql -{1\over 96} \cA_\mu{}^{ijkl}\cA^\mu{}_{ijkl}\,.
\end{equation}

The scalar potential in the $\cN=8$ theory is:
\begin{equation}\label{scalpot}
\cP\eql -g^2\big(\coeff{3}{4}\left|A_1{}^{ij}\right|^2-\coeff{1}{24}\left|A_{2i}{}^{jkl}\right|^2\big)\,,
\end{equation}
where 
\begin{equation}\label{Atensors}
A_1{}^{ij}\eql \coeff{4}{21} T_k{}^{ikj}\,,\qquad A_{2i}{}^{jkl}\eql -\coeff{4}{3}T_i{}^{[jkl]}\,,
\end{equation}
and
\begin{equation}\label{ttensor}
T_i{}^{jkl}~\equiv~\left(u^{kl}_{IJ}+v^{klIJ}\right)\left(u_{im}{}^{JK}u^{jm}{}_{KI}-v_{imJK}v^{jmKI}\right)\,,
\end{equation}
is the so-called  $T$-tensor.

\subsubsection{Quadratic fluctuations of the scalar action}
\label{QuadFlucts}

The simplest, and most canonical method of generating scalar fluctuations about a given scalar configuration is by using left multiplication of the $\rm E_{7(7)}$ matrices.  Let $\cV_0$ be a generic, constant $\rm E_{7(7)}$ matrix that represents some background scalar configuration. Fluctuations  around this background can be parametrized by another $\rm E_{7(7)}$ matrix of the form 
\begin{equation}\label{expandvar}
\cV(\phi)\eql \widetilde\cV(\phi)\,\cV_0~\equiv~ \exp\left(\sum_{i=1}^{70}\phi_iG_i\right)\,\cV_0\,,
\end{equation}
that depends upon space-time coordinates through the fluctuations, $\phi_i(x)$,  of the scalar fields. The sum in \eqref{expandvar} runs over all non-compact generators, $G_i$, of $\rm E_{7(7)}$ and the parametrization \eqref{expandvar} guarantees that the fields $\phi_i$  provide local coordinates on the coset at $\cV_0$. Furthermore, we have 
\begin{equation}
\label{pertdefn}
(\cD_\mu \cV)\, \cV^{-1}  ~=~(\cD_\mu \widetilde \cV)\,  \widetilde\cV^{-1}    \,.
\end{equation}

We consider $\widetilde\cV$ as a perturbation and expand it in the   scalar fluctuations
\begin{equation}\label{quadexpan}
\widetilde\cV\eql \mathbbmss{1}+\sum_{i=1}^{70}\phi_iG_i+{1\over 2}\sum_{i,j=1}^{70}\phi_i\phi_j G_iG_j+\ldots\,.
\end{equation}
Substituting this expansion in \eqref{pertdefn} and  then using \eqref{conndefn}  to the linear order to evaluate  \eqref{sckin}, yields  an elementary expression for the  kinetic terms  of the perturbation.  As a result one can always choose the basis of generators, $G_i$,  so as to obtain a canonical kinetic term  for the perturbing fields
\begin{equation}
e^{-1}\delta\cL_{\rm kin.}^{\mathcal{N}=8} \eql - {1 \over 2} \, \sum_{i=1}^{70}\, (\partial_\mu \phi_i)^2 \,.
\label{normkin}
\end{equation}
To determine the expansion of other terms in the action to quadratic order in scalar fluctuations, one also uses $\cV = \widetilde \cV  \cV_0$, but being careful to retain all terms in \eqref{quadexpan} up to and including quadratic order. 

Later in this paper  we will use this technique to calculate the spectrum of masses around critical points of the potential \eqref{scalpot} in the $\rm SU(3)$-invariant sector, for which $\cV_0$ will be the group element representing a critical point and  $\widetilde \cV$  some completely general quadratic order fluctuation in all 70 scalars. More generally, in Section~\ref{sufourcrit}, we will also consider a quadratic fluctuation expansion which is valid along an uncharged flow between two critical points of the potential.

\subsubsection{The SU(3)-invariant sector}

To define the $\rm SU(3)$-invariant sector it is useful to introduce Cartesian coordinates  $x_I$, $I=1,\ldots,8$, on $\RR^8$ and define the complex variables $z_1=x_1+ix_2\,,\ldots,z_4\eql x_7+ix_8$.  The $\rm SU(3)$ then acts on $(z_1,z_2, z_3)$  in the fundamental representation. From this one can then define two manifestly $\rm SU(3)$-invariant complex structures on  $\RR^8$ via:
 \begin{equation}
 \label{complstr}
J^\pm\eql {i\over 2}\left(\sum_{j=1}^3 dz_j\wedge d \bar  z_j\right)\pm {i\over 2}\,dz_4\wedge d\bar z_4\,,
\end{equation}
and  the corresponding two real four-forms
\begin{equation}\label{Fone}
F_1^+\eql J^+\wedge J^+\,,\qquad F_1^-\eql J^-\wedge J^-\,.
\end{equation}
The other real and manifestly $\rm SU(3)$-invariant forms are given by defining the complex four-forms:
\begin{equation}
\label{Ftwoth}
F_{23}^+ \eql dz_1\wedge dz_2\wedge dz_3\wedge dz_4\,,\qquad 
 F_{23}^- \eql dz_1\wedge dz_2\wedge dz_3\wedge d\bar z_4\,,
\end{equation}
and then taking the real and imaginary parts:
\begin{equation}
\label{}
F_2^\pm+iF_3^\pm\eql F_{23}^\pm\,.
\end{equation}
The $F_j^+$ are real and self-dual and the $F_j^-$ are real and anti-self-dual.  To get complex self-dual scalars $\Sigma_{IJKL}$ one must take real linear combinations of   $F_j^+$ and  $i F_j^-$ and these parametrize the manifold of $\rm SU(3)$-invariant scalar fields. 

As $E_{7(7)}$ generators, the $F_1^\pm$ commute with $F_2^\pm, F_3^\pm$ and in fact represent the non-compact generators of the coset
\begin{equation}
\label{cosetman}
\cM \eql \cM_{\rm SK}\times \cM_{\rm QK}~\equiv~  {\rm  {SU(1,1)\over U(1)}\times {SU(2,1)\over SU(2)\times U(1)}}\,.
\end{equation}
The first factor, parametrized by $F_1^\pm$, contains the two scalars of the vector multiplet and is a special K\"ahler manifold. The second factor, parametrized by  $F_2^\pm, F_3^\pm$,  is a quaternionic K\"ahler manifold and represents the scalars in the hypermultiplet. See Appendix \ref{appendixB} for a more extensive discussion of these coset spaces.

We parametrize  the scalar manifold, $\cM$,   in terms of three complex fields $\wom_1$ and $\wom_2$, $\wom_3$, 
by setting:\footnote{This parametrization differs slightly from the one  in \cite{Bobev:2009ms}, where we used
\begin{equation*}
\Sigma(w_1,w_2,w_3)\eql\sum_{j=1}^3 \left({\rm Re}(w_j)\,F_j^++i\,{\rm Im}(w_j)\,F_j^-\right)\,.
\end{equation*}
The relation between the two sets of fields is
\begin{equation*}
\wom_1\eql 2\, w_1\,,\qquad \wom_2\eql 2\left({\rm Re}(w_2)-i\,{\rm Re}(w_3)\right)\,,\qquad \wom_3\eql 2\left({\rm Im}(w_3)+i\,{\rm Im}(w_2)\right)\,.
\end{equation*}
} 
\begin{equation}
\label{esevengen}
\Sigma(\wom_1,\wom_2,\wom_3)  \eql  ~ {1\over 2}\left( {\rm Re}(\wom_1)\,F_1^++i\,{\rm Im}(\wom_1)\,F_1^-\right)  +{1\over 4}\left(\wom_2\,F_{23}^++{\rm c.c.}\right)+{1\over 4}\left(\wom_3\,F_{23}^--{\rm c.c.}\right)\,.
\end{equation}
These $\rm E_{7(7)}$  generators may be written in terms of the  $\rm SU(1,1)\times SU(2,1)$ Lie algebra generators as
\begin{align}
\label{fundgen}
\sigma_{(2)}(\wom_1) &\eql   \begin{pmatrix}
0 &   \wom_1 \\  \overline\wom_1  & 0 
\end{pmatrix}\,,
\qquad 
\sigma_{(3)}(\wom_2,\wom_3)  \eql  \begin{pmatrix}
0 & 0 & \wom_2 \\ 0 & 0 & \wom_3 \\
\overline\wom_2 & \overline\wom_3 & 0
\end{pmatrix}\,.
\end{align}

One can now exponentiate these matrices and introduce the standard projective coordinates, $z$ and $\zeta_1, \zeta_2$,  defined by
\begin{equation}
z  ~\equiv ~ {\wom_1 \tanh|\wom_1|\over |\wom_1|}\,,\qquad 
 \zeta_{i}  ~\equiv ~ {\wom_{i+1}\tanh (\sqrt{|\wom_2|^2+|\wom_3|^2})\over \sqrt{|\wom_2|^2+|\wom_3|^2}}\,,\quad i=1,2\,.
\end{equation}
Since $\omega_1$, $\omega_2$ and $\omega_3$ are arbitrary complex variables, the range of the new coordinates is $|z|<1$ and $|\zeta_1|^2+|\zeta_2|^2<1$.

In this parametrization, the $\rm SU(3)$-invariant truncation of the Lagrangian \eqref{sckin} yields the following:
\begin{equation}\label{kinetic}
e^{-1}\cL_{\rm kin.}\eql -g_{\zet\bzet}\,\partial_\mu \zet \partial^\mu\bzet-g_{\zeta_i\bzeta_j}\nabla_\mu \zeta_i\nabla^\mu\bzeta_j\,,
\end{equation}
which is the Lagrangian  of a gauged $\sigma$-model on the product coset space \eqref{cosetman} with the unique $\rm SU(1,1)$ and $\rm SU(2,1)$-invariant K\"ahler metrics
\begin{equation}\label{kahmetrics}
ds^2_{\rm SK}\eql 3\,{d\zet\,d\bzet\over (1-|z|^2)^2}\,,
\end{equation}
and
\begin{equation}\label{quatermetr}
ds^2_{\rm QK}\eql 
 {d\zeta _1 d\overline\zeta {}_1+d\zeta _2d\overline\zeta {}_2\over 1-|\zeta _1|^2-|\zeta _2|^2} +
{(\zeta _1d\overline\zeta {}_1+\zeta _2d\overline \zeta {}_2)(\overline \zeta {}_1d\zeta _1+\overline \zeta {}_2d\zeta _2)\over (1-|\zeta _1|^2-|\zeta _2|^2)^2}\,,
\end{equation}
respectively. 
The covariant derivative  of the charged scalar fields  is
\begin{equation}
\label{redcovder}
\nabla_\mu\zeta_i\eql \partial_\mu\zeta_i+ g \sum_{\alpha=0}^1 A_\mu^\alpha K_\alpha^{\zeta_i}\,,\qquad i=1,2\,,
\end{equation}
where $g$ is the coupling constant of the gauged supergravity, as in \eqref{covderu}, and the $K_\alpha$ are the Killing vectors on the quaternionic coset and will be defined in   \eqref{killvect} below.   Clearly, apart from the relative normalization of the two terms in the Lagrangian, $\cL_{\rm kin.}$ is completely fixed by the  $\rm SU(1,1)\times SU(2,1)$  global symmetry  and covariance.

\subsubsection{The scalar potential}
 
The scalar potential in the $\rm SU(3)$-invariant sector was first calculated in  \cite{Warner:1983vz} and more recently expressed using a complex superpotential in \cite{Ahn:2000mf} (see also \cite{Bobev:2009ms}). For the calculation of the potential, it is sufficient to work with four scalar fields in a gauge-fixed sector of the theory.  One can then recover  the dependence on all six fields by replacing gauge-fixed quantities with the corresponding  gauge invariants. The result we present here follows directly from the one in \cite{Bobev:2009ms}, after one corrects a typographical error in Eqs.\ (3.13) of that reference.  We have also directly verified the result presented here by  calculating it with all six fields present.\footnote{A direct calculation of the potential without gauge fixing has been also carried out recently in \cite{Ahn:2009as}.}

If one defines the manifestly gauge-invariant complex field by
\begin{equation}
\label{zetavar}
\zeta_{12} \eql {|\zeta_1|+ i\,|\zeta_2|\over 1+\sqrt{1-|\zeta_1|^2-|\zeta_2|^2} }\,,
\end{equation}
then the two  ``holomorphic''   superpotentials \cite{Ahn:2000mf}, obtained as eigenvalues of the $A_1$-tensor \eqref{Atensors} on the $\rm SU(3)$-invariant subspace, are:
\begin{equation}
\label{superpot}
{\cal W_+} \eql (1-|z|^2)^{-3/2}(1-|\zeta_{12}|^2)^{-2}\left[(1+z^3)\,(1+\zeta_{12}^4)+6\, z \, \zeta_{12}^2(1+z)\right]\,,
\end{equation}
and
\begin{equation}\label{superpotm}
{\cal W_-} \eql (1-|\zet |^2)^{-3/2}(1-|\zeta_{12}|^2)^{-2}\left[(1+\zet ^3)(1+\bar\zeta_{12}^4)+6\,\zet\, \bar\zeta_{12}^2(1+\zet )\right]\,.
\end{equation}
The remaining six eigenvalues of the $A_1$-tensor correspond to the six supersymmetries that are broken in the SU(3)-invariant sector.

The scalar potential \eqref{scalpot} on the $\rm SU(3)$-invariant fields is simply expressed in terms of either of the real superpotentials $\cW=|\cW_+|$ or $\cW=|\cW_-|$ (cf.\ \cite{Ahn:2000mf},\cite{Bobev:2009ms}):
\begin{equation}\label{potsupepot}
\cP\eql 2g^2\left[ \,4\, g^{\zet\bzet}\,{\partial \cW\over\partial \zet}{\partial \cW\over\partial \bzet}+4 \,g^{\zeta_i\bzeta_j}\,{\partial \cW\over\partial \zeta_i}{\partial \cW\over\partial \bzeta_{j}}-3\,\cW^2\,\right]\,,
\end{equation}
or equivalently,
\begin{equation}
\label{poten}
{\cal P}\eql  2\, g^2 \bigg[ {4 \over 3} (1 - |z|^2)^2 \left|{\partial | \cW_+ | \over \partial z}\right|^2  + (1 -
|\zeta_{12}|^2)^2  \left|{\partial  | \cW_+ | \over \partial \zeta_{12}} \right|^2\  -  3\,  |\cW_+|^2 \bigg] \,,
\end{equation}
or by a similar similar expression obtained by letting $\cW_+\rightarrow\cW_-$ and $\zeta_{12}\rightarrow -\ozeta_{12}$.
It should be clear that  once the explicit dependence on all six fields is unravelled, the actual  expression for the potential becomes quite involved.

\subsection{Gauge symmetries and vector fields}

\subsubsection{The embedding in  SO(8) }
 
The residual $\rm{U(1)} \times \rm{U(1)}$ gauge symmetry commuting with $\rm{SU(3)} \subset \rm{SO(8)}$ corresponds to phase rotations  $z_j\rightarrow e^{i\varphi}z_j $, $j=1,2,3$ and $z_4\rightarrow e^{i\psi}z_4$, respectively.  The two Abelian gauge fields, $A^\alpha$, $\alpha=0,1$, parametrize the (normalized) $\rm{SO(8)}$ generator represented by the block diagonal matrix
\begin{equation}\label{gaugefield}
(A^{IJ})\eql {\rm diag}\,
\left( \hbox{$ {1\over \sqrt 3}\left(\begin{matrix}
0 & A^1\\-A^1 & 0
\end{matrix}\right),{1\over \sqrt 3}\left(\begin{matrix}
0 & A^1\\-A^1 & 0
\end{matrix}\right),{1\over \sqrt 3}\left(\begin{matrix}
0 & A^1\\-A^1 & 0
\end{matrix}\right),\left(\begin{matrix}
0 & A^0\\-A^0 & 0
\end{matrix}\right) $}\right)\,.
\end{equation}

In the unitary gauge, one can identify the  $\rm U(1)\times U(1)$ gauge group with a subgroup of $\rm SU(2)\times U(1)\subset SU(2,1)$. In terms of the $\rm SU(2,1)$ generators, the gauge field \eqref{gaugefield} becomes
\begin{equation}
\label{ggaugefield}
\sum_{\alpha=0}^{1}A^\alpha T_\alpha\eql {i\over 4} \,
 \left(
\begin{matrix}
 A^0 + {1\over\sqrt 3} A^1 & 0 & 0\\[6pt]
  0 & - A^0 + {1\over \sqrt 3}A^1& 0\\[6 pt]
0 & 0 & -{2\over \sqrt 3}A^1
\end{matrix}
\right)\,.
\end{equation}

The Killing vectors, $K_\alpha$,   of the isometries  generated by the left action of the $T_\alpha$'s on the coset, can be easily read off from \eqref{fundgen} and \eqref{gaugefield}. They are
\begin{equation}\label{killvect}
K_0 \eql i\,\zeta_1 \partial_{\zeta_1}- i \,\zeta_2\partial_{\zeta_2}+{\rm c.c.}  \,,\qquad  K_1\eql\sqrt 3\,i\,\zeta_1 \partial_{\zeta_1}+ \sqrt 3 \,i \,\zeta_2\partial_{\zeta_2}+{\rm c.c.} 
\,.
\end{equation}

The gauge transformations of the scalar fields can be read off from the transformations of the corresponding four-forms,
\begin{equation}\label{gaugevarofsc}
z ~\longrightarrow~ z,\,\qquad \zeta_1~\longrightarrow~e^{i(3 \varphi+\psi)}\zeta_1\,,\qquad \zeta_2~\longrightarrow~e^{i(3 \varphi- \psi)}\zeta_2\,.
\end{equation}
Hence gauge fixing of the residual gauge symmetry amounts to fixing the phases of the scalar fields $\zeta_1$ and $\zeta_2$. 

\subsubsection{The  action}

The Lagrangian for the gauge fields in $\Neql 8$ four-dimensional supergravity is  \cite{de Wit:1982ig}:
\begin{equation}
\label{maxmax}
e^{-1}\cL_{gauge}^{\cN \eql  8} \eql -{1\over 8}\left[F_{\mu\nu\,IJ}^+(2 S^{IJ,KL}-\delta^{IJ}_{KL})F^{+\mu\nu}{}_{KL}+{\rm c.c.}\right]\,,
\end{equation}
where $F_{\mu\nu\,IJ}^+$ is the self-dual part of the field strength:
\begin{equation}
\label{fldstr}
F_{\mu\nu\,IJ}   \eql   \partial_\mu A_{\nu\,IJ}-g\,A_{\mu\,IK}A_{\nu\,KJ}-(\mu\leftrightarrow\nu)\,,
\end{equation}
and $S^{IJ,KL}\eql S^{KL,IJ}$ is a tensor defined in terms of the scalar $56$-beins:
\begin{equation}
\label{stens}
\left(u^{ij}{}_{IJ}+v^{ijIJ}\right)S^{IJ,KL}\eql u^{ij}{}_{KL}\,.
\end{equation}
The calculation of all components of this tensor is quite involved. However, all that is needed here are the components that couple to the $\rm{U(1)}\times \rm{U(1)}$ field strengths, and those depend only on the neutral scalar field, $z$. Hence one can solve \eqref{stens} for $S^{IJ,KL}$ with $\zeta_1$ and $\zeta_2$ set to zero, which considerably simplifies the algebra.

The final result for the truncated Lagrangian reads:
\begin{align}
\label{maxwell}
\begin{split}
e^{-1}\,\cL_{gauge} & \eql -{1\over 4} \sum_{\alpha,\beta=0}^1\left( \tau_{\alpha\beta}\, F_{\mu\nu}^{+ \alpha} F^{+\beta}{}^{\mu\nu}+\overline \tau_{\alpha\beta}\, F_{\mu\nu}^{- \alpha} F^{-\beta}{}^{\mu\nu}\right)\\[6pt]
&\eql -{1\over 4} \sum_{\alpha,\beta=0}^1 \left({\rm Re}(\tau_{\alpha\beta})F_{\mu\nu}^\alpha F^{\beta\, \mu\nu}+i\,{\rm Im}(\tau_{\alpha\beta})F_{\mu\nu}^\alpha\widetilde F^{\beta \, \mu\nu }\right)\,,
\end{split}
\end{align}
where
\begin{equation}
\label{}
F_{\mu\nu}^\alpha\eql \partial_\mu A_\nu^\alpha-\partial_\nu A_\mu^\alpha\,,
\end{equation}
and\footnote{We use $\eta^{\mu\nu\rho\sigma}=e^{-1}\epsilon^{\mu\nu\rho\sigma}$, $\epsilon^{0123}=1\,.$}
\begin{equation}
\label{fielst}
F_{\mu\nu}^{\pm\alpha}\eql {1\over 2}\left(F_{\mu\nu}^\alpha\pm\,\widetilde F_{\mu\nu}^\alpha
\right)\,,\qquad \widetilde F_{\mu\nu}^\alpha \eql {i\over 2}\eta_{\mu\nu}{}^{\rho\sigma} F_{\rho\sigma}^{ \alpha}\,,\qquad \alpha\eql 0,1\,.
\end{equation}
The symmetric tensor, $\tau_{\alpha\beta}$,  is given by:
\begin{align}
 \tau_{00} & \eql  {\left(1+2\bzet +3 \zet \bzet +3\zet ^2+2\zet ^3+\zet ^3\bzet \right)\over (1+\zet )^2\left (1-2\zet +2\bzet-\zet  \bzet \right)}\,,\\[6pt]
  \tau_{11}& \eql {\left(1-2\bzet +3 \zet \bzet +3\zet ^2-2\zet ^3+\zet ^3\bzet \right)\over (1+\zet )^2\left (1-2\zet +2\bzet-\zet  \bzet \right)}\,,\\[6pt]
 \tau_{01}&\eql \tau_{10}\eql \,-{2\sqrt{3}\,\zet (1+\zet \bzet )\over (1+\zet )^2\left (1-2\zet +2\bzet-\zet  \bzet \right)}\,.
 \label{tauans3}
\end{align}

This completes the description of the truncation of the bosonic action of the gauged $\Neql8$ supergravity theory down to that of the $\Neql2$ theory of interest. In Appendix \ref{appendixB} we will recast this action in a canonical form in terms of special K\"ahler and quaternionic K\"ahler geometry.

\section{The SU(4)$^-$  invariant sector}
\label{sufoursector}

We now further restrict our attention to the $\rm SU(4)^-$ invariant sector and demonstrate that the bosonic Lagrangian in this sector is precisely equivalent to the one used in \cite{Gauntlett:2009dn, Gubser:2009gp,Gauntlett:2009bh} to study certain non-supersymmetric RG-flows and symmetry breaking domain walls. Within this sector there are three non-trivial critical points, one of which was argued in \cite{Gauntlett:2009bh} to be the ground state for a holographic superconductor. Since we have obtained the theory from $\cN=8$ gauged supergravity we are then able to analyze the stability of this point outside of the $\rm SU(4)^-$ sector.  

\subsection{SU(4)$^-$ truncation}
\label{su4trunc}

The $\rm SU(4)^-$ is defined to be the subgroup of $\rm SO(8)$ that preserves the form, $F_{23}^-$. The complex structure $J^-$ is invariant under $\rm SU(4)^- \times U(1)$, where the $\rm U(1)$ is the remaining gauge symmetry. There are thus three $\rm SU(4)^-$ invariant scalar fields corresponding to the forms $F_j^-$, $j=1,2,3$, while the invariant gauge fields are obtained by arranging that (\ref{gaugefield}) be proportional to the matrix $J^-$ by setting
\begin{equation}
\label{vectr}
 A^0\eql -{1\over 2}B  \,,\qquad A^1\eql {\sqrt{3}\over 2} B\,,
\end{equation}
where $B$ is the new gauge field.  The scalar fields consist of one  pure imaginary  field, $z \equiv  i\eta$,  that is neutral under the $\rm U(1)$ and one charged complex scalar field, $\zeta_2\equiv \zeta$.   

It is interesting to note that this sector explicitly breaks the $\cN=2$ supersymmetry of the previous section. 
The supersymmetries and the gravitinos of the $\cN=8$ theory lie in the ${\bf 8_+}$, the spin-$1/2$ fermions lie in the ${\bf 56_+}$ of $\rm SO(8)$ and under $ \rm{SU(4)^-} \subset \rm{SO(8)}$ the ${\bf 8_+}$ branches to ${\bf 4}\oplus  {\bf {\overline4}}$ and the ${\bf 56_+}$ branches to $\mathbf{20} \oplus \mathbf{\overline{20}} \oplus\mathbf{4}\oplus\mathbf{\overline{4}} \oplus \mathbf{4}\oplus\mathbf{\overline{4}}$. In particular there are no singlets in these decompositions and so as a result, the $\rm SU(4)^-$ sector is purely bosonic.

The complete action that follows from the results in Section \ref{suthreesector} is:
\begin{equation}
\label{sufouraction}
\begin{split}
e^{-1}\cL   ~\eql~ & {1\over 2}R -{1\over 4}\tau(\eta)F_{\mu\nu}F^{\mu\nu}-{i\over 4}\theta(\eta)F_{\mu\nu}\widetilde F^{\mu\nu}- 3\,{\partial_\mu\eta\,\partial^\mu\eta\over (1-\eta^2)^2}-{\nabla_\mu\zeta\nabla^\mu\ozeta\over (1-|\zeta|^2)^2}-\cP_{\rm SU(4)}\,,
\end{split}
\end{equation}
where
\begin{equation}
\label{}
\tau(\eta)\eql {(1-\eta^2)^3\over 1+15\eta^2+15\eta^4+\eta^6}\,,\qquad \theta(\eta)\eql {2\eta(3+10\eta^2+3\eta^4)\over 1+15\eta^2+15\eta^4+\eta^6}\,,
\end{equation}
and
\begin{equation} 
\nabla_\mu\zeta\eql\partial_\mu\zeta-2igB_\mu\zeta\,.
\end{equation}
The scalar potential in this sector is
\begin{equation}\label{sufourpot}
\cP_{\rm SU(4)}\eql - 2 g^2 {(1+\eta^2)\,\big[\, 3(1-\eta^2)^2-4(1+\eta^2)^2|\zeta|^2\,\big]\over (1-\eta^2)^3(1-|\zeta|^2)^2}\,.
\end{equation}
It has three critical points  at:
(i) $\zeta=\eta=0$ with $\cP_*=-6g^2$,  (ii) $\eta=\sqrt 5-2\,,$ $\zeta=e^{i\alpha}/\sqrt 5$ with $\cP_*=-25\sqrt5/8\,g^2$, and (iii) $\eta=0$, $\zeta\eql e^{i\alpha}/\sqrt{2}$ with $\cP_*=-8g^2$. The first one is the maximally supersymmetric $\rm SO(8)$-invariant point of the $\cN=8$ theory. The second one is the $\rm SO(7)^-$ invariant point that is non-supersymmetric and is known to be unstable \cite{deWit:1983gs}.
The last one is the $\rm SU(4)^-$ invariant point and this also breaks all   supersymmetries. We will show below that this non-supersymmetric  point becomes unstable under scalar fluctuations in the full  $\cN=8$ theory.

\subsubsection{Comparison with \cite{Gauntlett:2009bh}}

It is worth pointing out that the bosonic Lagrangian, \eqref{sufouraction}, in the $\rm SU(4)^{-}$ sector of four-dimensional $\mathcal{N}=8$ gauged supergravity is exactly the same as the Lagrangian obtained in \cite{Gauntlett:2009zw,Gauntlett:2009bh}  from a consistent truncation of eleven-dimensional supergravity including massive modes.

The precise translation between the fields in \eqref{sufouraction} and the ones of Eq.\ (4.3) in \cite{Gauntlett:2009bh} is given by:
\begin{equation}\label{togaunt}
A_{{\rm GSW}} \eql \ds\frac{g}{2} B~, \qquad h_{{\rm GSW}} \eql {2\eta\over 1+\eta^2}~, \qquad \chi_{{\rm GSW}} \eql {2\over \sqrt 3}\,\zeta\,.
\end{equation}
The subscript ${\rm GSW}$ denotes the fields of \cite{Gauntlett:2009bh}, where $A_{{\rm GSW}}$ is the gauge field, $h_{{\rm GSW}}$ is the neutral scalar and $\chi_{{\rm GSW}}$ is the charged scalar. It is easy to check that with these field redefinitions the Lagrangian in Eq.\ (4.3) of \cite{Gauntlett:2009bh} matches with \eqref{sufouraction} provided one sets
\begin{equation}
\epsilon_{\rm GSW} = -1\,, \qquad \ds\frac{1}{16\pi G_{{\rm GSW}}} = \ds\frac{1}{2}\,,  \qquad  g^2=2\,,
\end{equation}
where $\epsilon_{\rm GSW}$ is the skew-whiffing parameter and $G_{{\rm GSW}}$ is the Newton's constant in \cite{Gauntlett:2009bh}, and $g$ is the coupling constant of gauged supergravity in \eqref{covderu}.

\subsection{The instability of the SU(4)$^-$ critical point}
\label{sufourcrit}

The $\rm SU(4)^-$ critical point, and the uncharged flow to it, only involves the scalar field, $\zeta$, and the neutral scalar, $\eta$, can be consistently set to zero.  The residual $\rm U(1)$ symmetry means that the potential \eqref{sufourpot} only depends upon $|\zeta|$ and we choose a gauge such that $\zeta$ is {\it real}. For $\eta=0$, the potential collapses to
\begin{equation}\label{simppoten}
\widehat\cP\eql -2 g^2 \,(3-4|\zeta|^2)(1-|\zeta|^2)^{-2}\,.
\end{equation}

The  70 scalar fluctuations, $\phi_i$, at any point $\zeta$ comprise of 35 pseudoscalars and 35 scalars of $\cN  =  8$ supergravity, in $\mathbf{35_{-}}$ and $\mathbf{35_+}$ of $\rm SO(8)$, respectively. The former  correspond  to ``flux modes'' in eleven dimensions,  and they break down into the $\rm SU(4)^-$ representations:\footnote{See Appendix \ref{appendixC} for more details on branching rules for various $\rm SO(8)$ representations.} $ {\bf 20'} \oplus {\bf 6}  \oplus {\bf 6} \oplus {\bf 1} \oplus {\bf 1} \oplus{\bf 1} $. The latter correspond to ``metric modes'' in eleven dimensions,  and they break down into the $\rm SU(4)^-$ representations: $ {\bf 15} \oplus {\bf 10} \oplus{\bf \overline {10}}$. 
We have applied the technique outlined in Section \ref{QuadFlucts} to expand the scalar Lagrangian to quadratic order and to obtain the spectrum of scalar fluctuations, $\phi_i$, and we find
\begin{equation}\label{exppot}
\begin{split}
\cP  &\eql  ~  \widehat \cP + \delta|\zeta|{d\over d|\zeta|}\widehat \cP \\ 
&  + \,^{(1)}\mu_{\bf 1}\, \phi_1^2 + \,^{(2)}\mu_{\bf 1}\, \phi_2^2+\,^{(3)}\mu_{\bf 1}\, \phi_3^2 +
\mu_{\bf 6\oplus \overline 6}\, \sum_{i=4}^{15}\, \phi_i^2 + \mu_{\bf 20'}\,\sum_{i=16}^{35}\, \phi_i^2\\
& + \mu_{\bf 10\oplus \overline {10}} \sum_{i=36}^{55}\, \phi_i^2  + \mu_{\bf 15} \,  \sum_{i=56}^{70}\, \phi_i^2 \,,
\end{split}
\end{equation}
where  $\widehat \cP$ is the background potential \eqref{simppoten}. The coefficients of the quadratic terms in the expansion are:
\begin{equation}\label{deffluct}
\begin{split}
^{(1)}\mu_{\bf 1} & \eql  -2g^2 \,(1-3 |\zeta|^2-2|\zeta|^4)(1-|\zeta|^2)^{-2}\,,\\[6 pt]
^{(2)}\mu_{\bf 1} & \eql -2g^2\,(1-4 |\zeta|^2)(1-|\zeta|^2)^{-2}\,,\\[6 pt]
^{(3)}\mu_{\bf 1} & \eql -2g^2\,(1+|\zeta|^2)(1- 2|\zeta|^2)(1-|\zeta|^2)^{-2}\,,\\[6 pt]
\mu_{\bf 6\oplus \overline 6}& \eql -2g^2\,(1- 2|\zeta|^2)(1-|\zeta|^2)^{-2}\,,\\[6 pt]
\mu_{\bf 20'} & \eql - 2g^2 (1-|\zeta|^2)^{-1}\,,\\[6 pt]
\mu_{\bf 10\oplus \overline {10}} & \eql -g^2\, (2-3|\zeta|^2-|\zeta|^4)(1-|\zeta|^2)^{-2}\,,\\[6 pt]
\mu_{\bf 15} &\eql  -2g^2\,(1- 2|\zeta|^2)(1-|\zeta|^2)^{-2}\, .
\end{split}
\end{equation}
The scalar fluctuations, $\phi_i$, have been normalized to have kinetic terms (\ref{normkin}) and the linear mixing in (\ref{exppot}) is due to the fact that $ \phi_1= \sqrt 2\,(1-|\zeta|^2)^{-1} \delta|\zeta|$.  Note that this shows that setting all the scalars  except real $\zeta$ to zero is indeed a consistent truncation within the $\rm SU(4)^-$ sector. We will be mostly interested in the expressions \eqref{deffluct} evaluated at the two fixed points but we would like to point out that they are valid along an uncharged supergravity domain wall connecting the two points.

At a critical point of the potential, $\mathcal{P} = \mathcal{P}_*$, the dimensionless mass for a scalar field in $AdS_4$ of radius $L$ is
\begin{equation}
m^2_{{\bf r}} \equiv M^2_{{\bf r}} L^2 = - 6\, \ds\frac{ \mu_{{\bf r}}}{\mathcal{P}_*}\,,
\end{equation}
where ${\bf r}$ is one of the representations above. The stability condition  (\ref{Bfcrit}) at a critical point then  becomes %
\begin{equation}
 m^2_{{\bf r}} \geq - \ds\frac{9}{4} \,.
 \label{stabcond} 
\end{equation}

At at the $\rm SO(8)$ point,  $\zeta=0$, one has $\cP_* = -6g^2$, $L^2 = 1/(2g^2)$ and  $m_{{\bf r}}^2 = -2 $, for all the masses. As one expects from supersymmetry, this point is stable.

At the $\rm SU(4)^-$ point,  $\zeta=1/\sqrt 2$, one has $ \cP_*  = -8 g^2$, $L^2 = 3/(8g^2)$ and 
\begin{equation}
\label{SU4masses}
^{(1)}m_{\bf 1}^2\eql \,^{(2)}m_{\bf 1}^2 \eql 6 \,,  \quad  ^{(3)}m_{\bf 1}^2 \eql m^2_{\bf 6\oplus \overline 6}\eql m^2_{\bf 15 }\eql  0 \,, \quad m^2_{\bf 10\oplus \overline {10}}\eql  -\ds\frac{3}{4}\,, \quad   m^2_{\bf 20'}\eql -3    \,.
\end{equation}
Thus scalar fluctuations in the ${\bf 20'}$ of $\rm SU(4)^-$ with mass, $m_{\bf 20'}$, are  {\it unstable\/} in $\Neql\,8$ supergravity.

\subsection{Some comments on the unstable modes}

The first thing we wish to note is that under the $\rm SU(3)$
subgroup, the ${\bf 20'}$ of $\rm SU(4)^-$ decomposes as ${\bf 8} \oplus
{\bf 6} \oplus {\bf \bar 6}$ and, most significantly, it does not
contain any $\rm SU(3)$ singlets.   This means that the instability is
{\it completely invisible\/} from the perspective of the $\Neql\,2$
supergravity theory that could be used to derive the existence
of the fixed point in the first place.  The instability thus comes
from {\it other\/} supergravity modes that arise through the embedding
into $\mathcal{N}=8$ supergravity. This observation is interesting in that it
demonstrates an important issue  in using low-dimensional effective or
truncated field theories to study non-supersymmetric vacua. There may
be hidden pathologies in the truncated theory and these pathologies
only become evident once one properly embeds the theory into a proven
holographic framework.

It is tempting to consider the possibility that these unstable modes
may drive one away from the $\rm SU(4)^-$ point to a supersymmetric
critical point elsewhere and thus restore stability. This would have been a
highly desirable circumstance since it certainly could restore some
control over the holographic superconductors which have been studied
in this sector. However, the known {\it supersymmetric} critical
points involve non-trivial values for scalars in both the $\bf 35_+$
{\it and} the $\bf 35_-$ and the reasons for this are clear from the
perspective of the dual UV field theory: the $\bf 35_-$ corresponds to
the fermion mass matrix while the $\bf 35_+$ corresponds to the boson
mass matrix and to preserve supersymmetry there must be some equality
between the fermion and boson masses.

From the scalar spectrum at the $\rm SO(7)^-$ and $\rm SU(4)^-$ points
presented in Appendix \ref{appendixA}, we see that the instabilities at each point correspond to the
modes that were dual in the UV to fermion masses. Specifically, at
the $\rm SO(7)^-$ point, there is a ${\bf 27}\subset {\bf 35_-}$
worth of unstable modes, which corresponds to a $7\times 7$ fermion
mass matrix, while at the $\rm SU(4)^-$ point the unstable modes
transform in the ${\bf 20'}\subset {\bf 35_-}$ which is dual to a
$6\times 6$ fermion mass matrix. This indicates that the unstable
modes will {\it not} drive the system to a supersymmetry restoring
vacuum. It is interesting to note from Table \ref{criticaltable} that
the $\rm SU(4)^-$ critical point has the lowest cosmological constant
in the $\rm SU(3)$-invariant sector, so there cannot be a flow from
it to any known supersymmetric vacuum.\footnote{Note that in the list of new critical points found in \cite{Fischbacher:2009cj} there is a critical point with a smaller cosmological constant which may be supersymmetric. It would be quite interesting to study this point and the possible flows to it further.}

However, to properly resolve this issue, one must really perform an
analysis of the actual holographic RG flow or symmetry breaking domain wall, not just the critical
point as we have done here. In our earlier work  \cite{Bobev:2009ms}
we found a certain universality in the space of holographic RG flows
within the $\rm SU(3)$-invariant sector, namely that there is a one-parameter
family of mass deformations that generically preserve only $\rm
SU(3)$ symmetry and $\cN=1$ supersymmetry. Apart from one
special flow in this family, which preserves a $\rm G_2$ subgroup of $\rm SO(8)$, the flows
terminate at the $\rm SU(3)\times U(1)$-invariant $\cN=2$
supersymmetric fixed point. It is conceivable that there might be a
similar universality in the space of non-supersymmetric RG flows. Specifically, it would be interesting to determine the IR fate of the general class
of RG flows driven by the following mass deformation\footnote{We are suppressing
the complications of monopole operators when adding a mass deformation to the ABJM
theory \cite{Aharony:2008ug} by taking the liberty of working in the
BLG theory \cite{Bagger:2007jr, Gustavsson:2007vu}.} in the $\rm
SU(3)$-invariant sector of the dual field theory,
\be
\delta\cL_{CFT} = m_1 \lambda_7^2 + m_2 \lambda_8^2 + m^2_3 \Phi_7^2 + m^2_4 \Phi_8^2\,,
\ee
where $\lambda_7$ and $\lambda_8$ are two fermions and $\Phi_7$ and $\Phi_8$ are two bosons and  $m_a$ are the four mass parameters. Several related issues regarding supersymmetric flows with non-trivial
profiles for the gauge field have been addressed recently in the type IIB
context in \cite{Bobev:2010de}.

\section{The instability in eleven dimensions}
\label{sec:instability}

\subsection{A brief outline}

Having established the instability of the $\rm SU(4)^-$ critical point at the level of four-dimensional $\Neql 8$ gauged supergravity, we now show that the corresponding solution  in M-theory is also unstable.  We do this by linearizing  the equations of motion of eleven-dimensional supergravity:%
\footnote{We use conventions with the mostly plus metric and $\epsilon^{01\ldots10}=+1$. The reader may note that the sign of the $\epsilon$-symbol is opposite to that   in \cite{Gauntlett:2002fz,Gauntlett:2009bh}.  Our  four-form flux, $F_{(4)}$, is normalized such that   $F_{(4)}= -{ 1\over 2} G_{(4)}$, where $G_{(4)}$ is the flux in \cite{Gauntlett:2002fz,Gauntlett:2009bh}. Then  the second equation \eqref{Mth:maxeqs} written in terms of forms is $d*F_{(4)}+F_{(4)}\wedge F_{(4)}=0$.  } 
\begin{equation}\label{Mth:eineqs}
R_{MN}+g_{MN}R\eql {1\over 3} F_{MPQR}F_{N}{}^{PQR}\,,
\end{equation}
\begin{equation}\label{Mth:maxeqs}
\nabla_MF^{MNPQ}\eql -{1\over 576}{1\over\sqrt{-g}}\epsilon^{NPQR_1\ldots R_8}F_{R_1\ldots R_4}F_{R_5\ldots R_8}\,,
\end{equation}
around the PW solution  \cite{Pope:1984bd,Pope:1984jj},  which is the uplift of the $\rm SU(4)^-$ critical point of the four-dimensional theory. 
The PW background is a product  space,  $AdS_4\times M_{7}$, where the internal manifold, $M_7$, is the sphere, $S^7$,  with a deformed $\rm SU(4)$-invariant metric. The unstable modes can be constructed using the harmonic expansion of the metric and of the flux around the background configuration, $(\circg_{MN},\circF_{MNPQ})$:
\begin{align}\label{espansion}
g_{MN}  & \eql\circg_{MN}+\sum_\Lambda  \,\phi^{(\Lambda)}\,H_{MN}^{(\Lambda)}\,,\\[6 pt]
\label{expandflux}
F_{MNPQ}  & \eql\circF_{MNPQ}+ \sum_\Lambda \,\phi^{(\Lambda)}\, \Phi_{MNPQ}^{(\Lambda)}+\ldots \,,
\end{align}
where $\phi^{(\Lambda)}(x)$ are the scalar fields   on $AdS_4$ and the omitted terms in \eqref{expandflux} contain derivatives of those fields. The metric harmonics,  $H^{(\Lambda)}_{MN}$, and the flux harmonics,   $\Phi_{MNPQ}^{(\Lambda)}$,   have components only along the internal manifold and transform in irreducible representations of the $\rm SU(4)$ symmetry group of the background. 

We  show explicitly that there is a consistent choice of  harmonics in the expansions \eqref{espansion} and \eqref{expandflux} in a $\bf 20'$ representation of $\rm SU(4)$ such that the linearized equations  \eqref{Mth:eineqs} and \eqref{Mth:maxeqs}, together with the Bianchi identity, $dF_{(4)}=0$, reduce to 
\begin{equation}\label{unstableeqs}
\Big(\Box_{AdS_4}+{3\over L^2}\Big)\,\phi^{(\lambda)}(x)\eql 0\,,\qquad \lambda=1,\ldots,20\,.
\end{equation}
This proves the instability of the background in eleven dimensions, cf. \eqref{SU4masses}. 

Furthermore, using the exact uplift formula for the metric, which relates scalar fields of  $\cN=8$ gauged supergravity and the   metric in eleven-dimensional supergravity at the full non-linear level \cite{deWit:1984nz}, we  show that the unstable modes above are precisely the    unstable modes in four dimensions derived   in Section \ref{sufourcrit}. 

While many technical details of our analysis rely  on the high degree of symmetry and the resulting simplification of the harmonic expansions, it is clear that the method is quite robust and similar results might hold for more general backgrounds in which $S^7$ in the PW solution is replaced with a Sasaki-Einstein manifold, $\rm SE_7$. We discuss this possibility at the end of the section.

\subsection{The Pope-Warner solution}
\label{popewarnersol}

The original PW solution \cite{Pope:1984bd,Pope:1984jj} of eleven-dimensional  supergravity  that we analyze here is constructed in terms of geometric data associated with the Hopf fibration of $S^7$ over $\CC\PP^3$, which determines both the  metric on the internal manifold  and the background flux. The eleven-dimensional metric and the flux are given by: 
\begin{equation}\label{popewarmet}
ds^2_{11}\eql ds^2_{AdS_4}+{8\over 3}L^2\left[\,ds^2_{\CC\PP^3}+2(d\psi+A)^2\,\right]\,,
\end{equation}
and\footnote{The first term in the flux has a sign which in our conventions corresponds to a ``skew-whiffed" Freund-Rubin solution. This sign is uniquely determined for the PW solutions \cite{Pope:1984bd,Pope:1984jj}.}
\begin{equation}\label{fluxpw}
F_{(4)}\eql {\sqrt 3\over 2L} {\rm vol}_{AdS_4} +\sqrt{32\over 27}\,L^3 d \left( e^{4i\psi}K+{\rm c.c.}\right)\,,
\end{equation}
where $ds_{AdS_4}^2$ is the metric on $AdS_4$ of radius $L$,   $ds^2_{\CC\PP^3}$ is the Fubini-Study metric, $A$ is the potential for the corresponding  K\"ahler form,  $\psi$ is the angle along the Hopf fiber, and $K$ is a holomorphic $(3,0)$ form on $\CC\PP^3$. The internal metric is normalized such that $ds^2_{\CC\PP^3}+(d\psi+A)^2$ is the $\rm SO(8)$-invariant metric on  $S^7$ with unit radius. The rescaling of the internal metric in \eqref{popewarmet} along the Hopf fiber breaks the isometry   to $\rm SU(4)\times U(1)$, with only $\rm SU(4)$ preserved by  the   flux \eqref{fluxpw}. 

In terms of   standard complex coordinates on $\CC\PP^3$ the Fubini-Study metric is
\begin{equation}\label{FSmetric}
ds^2_{\CC\PP^3}\eql {d z_id\bzet_i\over (1+\sum_k|\zet_k|^2)}-{(\zet_id\bzet_i)(\bzet_jd\zet_j)\over (1+\sum_k |\zet_k|^2)^2}\,,
\end{equation}
while the K\"ahler potential, $A$, and the holomorphic 3-form, $K$, are given by
\begin{equation}\label{Kpot}
A\eql {i\over 2}{\zet_j d\bzet_j-\bzet_jd\zet_j\over (1+\sum_k |\zet_k|^2)}\,,
\end{equation}
and
\begin{equation}\label{omegaform}
K\eql - {dz_1\wedge dz_2\wedge dz_3\over (1+\sum_k|\zet_k|^2)^{2}}\,.
\end{equation}
Using the identity $dK= 4 \,i\, A\wedge K$ it is straightforward to verify that the internal part of the flux \eqref{fluxpw} is given by
\begin{equation}\label{intflux}
F_{(4)}^{\rm int}\eql {16\over 3}\sqrt{{2\over 3}}\,i\, L^3 (\vartheta\wedge\Omega -\vartheta\wedge\overline \Omega)\,,
\end{equation}
where $\vartheta=d\psi+A$ is the 1-form along the Hopf fiber, which is also dual to the Killing vector $ \partial_\psi$, and $\Omega\eql e^{4i\psi}K$. 

The forms $\vartheta$ and $\Omega$,  have a simple realization  in terms of $\CC^4$, with complex coordinates, $\xi^a$, $a=1,\ldots,4$,  in which $S^7$ is embedded at unit radius by setting
\begin{equation}\label{hopfcoor}
\zxi^j\eql {z_j e^{i\psi}\over (1+\sum_k|z_k|^2)^{1/2}}\,,\qquad  \zxi^4\eql {e^{i\psi}\over (1+\sum_k|z_k|^2)^{1/2}}\,,\qquad   j=1,2,3\,.
\end{equation}
Then $\vartheta$ is the pullback onto $S^7$ of the form\footnote{We denote complex conjugate coordinates, $\overline {\xi^a}$, by $\xi^{\bar a}$. The indices in $\mathbb{C}^4$ are raised and lowered 
using the complex structure, $J_{a\overline b}=i\,\delta_{a\bar b}$, such that $\xi_{ a}=J_{a\bar b}\,\xi^{\bar b}$, etc.}
\begin{equation}\label{etaform}
\vartheta\eql -{1\over 2}\left( \xi_ad\xi^a+\xi_{\bar a}d\xi^{\bar a}\right)\,,
\end{equation}
and 
\begin{equation}\label{theomegaz}
\Omega\eql {1\over 6} \,\epsilon_{abcd}\, \zxi^a\,d\zxi^b\wedge d\zxi^c\wedge d\zxi^d\,,
\end{equation}
is the potential for the canonical holomorphic $(4,0)$ form $d\xi^1\wedge d\xi^2\wedge d\xi^3\wedge d\xi^4$ in $\CC^4$. This form is invariant under $\rm SU(4)$ and  defines it as a subgroup of $\rm SO(8)$. Finally the K\"ahler form
\begin{equation}\label{kahlerform}
J\eql J_{a\bar b}\,d \xi^a\wedge  d\xi^{\bar b}\,,
\end{equation}
descends to the K\"ahler form on $\CC\PP^3$.

\subsection{The linearized solution for the unstable modes}
\label{linearmodes}

Since the internal manifold of the PW solution is a deformed sphere, a linearized perturbation around this background can be analyzed systematically using  harmonic expansion on $S^7$,\footnote{An extensive discussion of   harmonic  expansions in the context of supergravity can be found in  \cite{Duff:1986hr,Castellani:1991et,PvanNreview:2008}.} in which the $\rm SO(8)$ harmonics are decomposed into irreducible $\rm SU(4)$ components. 

From the analysis in  Section \ref{sufourcrit} we know that  the unstable modes transform in   $\bf 20'$   of $\rm SU(4)$, which arises in the branching  of  $\bf 35_-$ of $\rm SO(8)$. It is a standard result for the  Kaluza-Klein spectrum  around the Freund-Rubin solution (see   \cite{Duff:1986hr} and   references therein), which corresponds to the $\rm SO(8)$  critical point of $\cN=8$ gauged supergravity, that at the linearized level the 35 pseudoscalar fields uplift   
to pure flux modes with the harmonic expansion of the form,
\begin{equation}\label{linfluxall}
P_{ABCD}(x)\,d{\cal K}^{AB}\wedge d{\cal K}^{CD}\,,
\end{equation}
where  $P_{ABCD}(x)$ are components of a real self-dual tensor in $\RR^8$ and ${\cal K}^{AB}$ are one forms dual to the $\rm SO(8)$ Killing vectors on $S^7$. 

The identification of the scalar fields with the flux and metric modes at the $\rm SU(4)^-$ critical point is obscured by the background flux, but we still expect that the $\bf 20'$ component in the decomposition of the \eqref{linfluxall} harmonic under $\rm SU(4)$  provides the leading term in the expansion of the unstable modes in the flux. Those components are spanned by the {\it real\/} harmonics,
\begin{equation}\label{flxharm}
 \Phi^{(\lambda)}_{(2,2)}\eql {1\over 4}\,\Phi^{(\lambda)}{}_{ab\bar c\bar d}\,d\xi^a\wedge d\xi^b\wedge d \xi^{\bar c} \wedge d \xi^{\bar d}\,,\qquad \lambda=1,\ldots\,,20\,,
\end{equation}
where $\Phi^{(\lambda)}{}_{ab\bar c\bar d}$ are components of a constant complex tensor, $\Phi^{(\lambda)}{}_{ab\bar c\bar d}=(\Phi^{(\lambda)}{}_{cd\bar a\bar b})^*$, that is antisymmetric in $[ab]$ and $[cd]$ and $J$-traceless. Hence   $\Phi^{(\lambda)}_{(2,2)}$ is  a primitive $(2,2)$ form of degree four in $\CC^4$.  
 
It is clear that in the presence of the internal background flux \eqref{intflux},  flux fluctuations \eqref{expandflux} with   harmonics of the form \eqref{flxharm} will give rise to a non-vanishing linearized energy-momentum tensor in the Einstein equation \eqref{Mth:eineqs}, and hence source non-vanishing metric fluctuations. This is a new feature in comparison with the $\rm SO(8)$ point, but not entirely unexpected. Indeed, similar metric fluctuations  arise in the analysis in \cite{Biran:1984jr,Page:1984fu} of the unstable modes around the Englert background \cite{Englert:1982vs}, that also has a non-vanishing internal flux. 

Since the coupling between the flux and the metric modes arises from the internal flux, which for the PW solution is essentially the (anti-)holomorphic volume form in $\CC^4$, the obvious  metric harmonics to consider are given in terms of the flux harmonics \eqref{flxharm}
\begin{equation}\label{fluxtomethar}
H^{(\lambda)}{}_{abcd}\eql {1\over 2}\,\epsilon_{cd}{}^{\bar e\bar g}\,\Phi^{(\lambda)}{}_{ab\bar e\bar g}\,,
\end{equation}
and their complex conjugate. One can verify that those constant tensors satisfy
\begin{equation}
H^{(\lambda)}_{abcd}\eql H^{(\lambda)}_{[ab][cd]}\eql H^{(\lambda)}_{[cd][ab]}\,, \qquad H^{(\lambda)}_{[abcd]}\eql 0\,,
\end{equation}
and, modulo the reality condition, are completely determined by these properties.

Our Ansatz for the  fluctuations of  the metric is given in terms of the harmonics \eqref{fluxtomethar}
\begin{equation}\label{metrhar}
\delta g_{z_iz_j}dz_idz_j\eql {1\over 4} \sum_{\lambda=1}^{20}\phi^{(\lambda)}(x)\,H^{(\lambda)}{}_{abcd}\,\xi^a\xi^cd\xi^bd\xi^d\,,
\end{equation}
and similarly for the complex conjugate components. A direct calculation of the deformed metric using exact uplift formulae shows that this Ansatz is in fact complete. We will return to this more technical discussion in Section \ref{metricmodes}
 below. Here let us note that the deformation is  only for the $(2,0)$ and $(0,2)$ components of the metric along $\CC\PP^3$ and that neither the warp factor nor the Hopf-fibered structure of the background metric are deformed by the fluctuations. 

By comparing \eqref{espansion} with \eqref{metrhar} one can read off the harmonics, $H_{MN}^{(\lambda)}$, in the expansion \eqref{espansion}. Using it to compute the linearized left hand side of the Einstein equations 
\eqref{Mth:eineqs}  yields a diagonal expansion
\begin{equation}\label{leftpert}
\delta E_{MN}\eql -{1\over 2}\sum_{\lambda=1}^{20}\,\left (\Box_{AdS}-{9\over L^2}\right)\phi^{(\lambda)}(x )\,H^{(\lambda)}_{MN}\,,
\end{equation}
where $E_{MN}$ stands for the combination of the Ricci tensor and the Ricci scalar in \eqref{Mth:eineqs}.

The full Ansatz for the fluctuations of the four-form flux turns out to be more complicated in that in addition to  the original $(2,2)$ forms in  \eqref{flxharm} one must introduce  $(2,2)$, $(3,1)$ and $(1,3)$  forms of degree six   that are sourced by  metric fluctuations 
 in the Maxwell equations \eqref{Mth:maxeqs}. 

The holomorphic structure of the metric fluctuations \eqref{metrhar} implies that there is no correction due to the metric modes to the right hand side of the Maxwell equations. However, there are terms on the left hand side that involve the fluctuations of the metric contracted  with the background flux
\begin{equation}\label{upfluxo}
\delta\circF{}^{MNPQ}\eql \delta g^{MM'}\,\circg{}^{NN'}\circg{}^{PP'}\circg{}^{QQ'}\circF_{M'N'P'Q'}+\ldots \,.
\end{equation}
Cancellation of those terms in some Maxwell equations requires additional $(2,2)$ factors in the flux, which in turn yield  non-vanishing contributions to the right hand side of other equations. By a systematic analysis of all Maxwell equations we have determined that all additional harmonics in the flux fluctuations can be read off from those that are present  in the harmonic expansion of the three-form $\nabla^M\delta\circF_{MNPQ}$, or, equivalently, its exterior derivative,  and arise from the following  four-forms in $\CC^4$:
\begin{equation}\label{firstform}
\Phi^{(\lambda)}_{(3,1)}\eql {1\over 2}\,\Phi^{(\lambda)}{}_{ab\bar c\bar d}\,J_{e\bar g}\,\xi^a\xi^e\,d\xi^b\wedge d\xi^{\bar c}\wedge d\xi^{\bar d}\wedge d\xi^{\bar g}\,,
\end{equation}
\begin{equation}\label{thirdform}
\Phi^{(\lambda)}_{(1,3)}\eql - {1\over 2}\,\Phi^{(\lambda)}{}_{ab\bar c\bar d}\,J_{e\bar g}\,\xi^{\bar c}\xi^{\bar g}\,d\xi^a\wedge d\xi^b\wedge d\xi^e\wedge d\xi^{\bar d}\,,
\end{equation}
and
\begin{equation}\label{scndform}
\begin{split}
\widetilde\Phi{}^{(\lambda)}_{(2,2)}& \eql  \Phi^{(\lambda)}{}_{ab\bar c\bar d}\,J_{e\bar g}\,
\xi^a\xi^{\bar c}\,d\xi^{b}\wedge d\xi^{e}\wedge d\xi^{\bar d}\wedge d\xi^{\bar g} \\[6 pt]&
 -{1\over 2}  \Phi^{(\lambda)}{}_{ab\bar c\bar d}\,J_{e\bar g}\, \left[\xi^e\xi^{\bar c}\,d\xi^{a}\wedge d\xi^{b}\wedge d\xi^{\bar d}\wedge d\xi^{\bar g}
+ \xi^a\xi^{\bar g}\,d\xi^{b}\wedge d\xi^{e}\wedge d\xi^{\bar c}\wedge d\xi^{\bar d} \, \right]\,.
\end{split}
\end{equation}
Hence we have two four-forms: the original $(2,2)$ form, $\Phi^{(\lambda)}_{(2,2)}$, defined in \eqref{flxharm} and
\begin{equation}\label{secondflux}
\Phi{}^{(\lambda)}_{(4)}~\equiv~ \Phi^{(\lambda)}_{(3,1)}+\widetilde\Phi{}^{(\lambda)}_{(2,2)}+
\Phi^{(\lambda)}_{(1,3)}\,.
\end{equation}
One verifies that both are closed
\begin{equation}\label{closure}
d\Phi^{(\lambda)}_{(2,2)} ~\equiv~d\Phi{}^{(\lambda)}_{(4)}\eql0\,.
\end{equation}

Let $C_{(2,1)}^{(\lambda)}+C_{(1,2)}^{(\lambda)}$, where
\begin{equation}\label{potentialsone}
C^{(\lambda)}_{(1,2)}\eql {1\over 8}\,\Phi^{(\lambda)}{}_{ab\bar c\bar d}\,\xi^{ a}\wedge d\xi^{b }\wedge d \xi^{\bar c} \wedge d \xi^{\bar d}\,,\end{equation}
and $C_{(2,1)}^{(\lambda)}$ is the complex conjugate, be the  potential for $\Phi^{(\lambda)}_{(2,2)}$,
\begin{equation}\label{potentials}
d\big(C_{(1,2)}^{(\lambda)}+C_{(2,1)}^{(\lambda)}\big)\eql \Phi^{(\lambda)}_{(2,2)}\,,
\end{equation}
and let similarly $C_{(3)}^{(\lambda)}$ be the potential for $\Phi{}^{(\lambda)}_{(4)}$.%
\footnote{One can check that the potential $C_{(3)}^{(\lambda)}$ is proportional to the three-form $\nabla^M\delta\circF_{MNPQ}$.}
We now consider the following fluctuations of the four-form flux around the PW background:
\begin{equation}\label{realflcu}
\delta F_{(4)}\eql  \sum_{\lambda=1}^{20} d \left[ \phi^{(\lambda)}(x)\Big(
\kappa_{(2,2)} \big(C_{(1,2)}^{(\lambda)}+C_{(2,1)}^{(\lambda)}\big)+\kappa_{(4)}C_{(3)}^{(\lambda)} \Big)\right]\,.
\end{equation}

By expanding the equations of motion to the linear order in the metric fluctuations \eqref{metrhar} and the flux fluctuations \eqref{realflcu} above we have verified explicitly that for
\begin{equation}\label{kappas}
\kappa_{(2,2)}\eql {L\over\sqrt 6} \,,\qquad \kappa_{(4)}\eql -{i\over 2\sqrt 6}\,L\,,
\end{equation}
the linearized equations reduce to \eqref{unstableeqs}.

\subsection{Unstable metric modes from a consistent truncation Ansatz}
\label{consisttranz}

\subsubsection{From four to eleven dimensions}
\label{fourelev}

It is widely believed but difficult to prove, that $\cN=8$ gauged supergravity in four dimensions is a  Kaluza-Klein reduction of the $\cN=1$ supergravity in eleven dimensions on $S^7$, such that   the full non-linear lower dimensional theory can be viewed as a consistent truncation of the higher dimensional one. In particular, it entails  that any solution to the field equations in four dimensions can be uplifted to a solution in eleven dimensions. 

This relation has been made explicit in various subsectors of the theory,%
\footnote{See, for example, \cite{deWit:1984nz,{de Wit:1986iy}} for earlier work and \cite{Gauntlett:2009zw,{Gauntlett:2009bh}} for some recent results that are pertinent to the discussion here.}
in particular, the eleven-dimensional  general metric corresponding to a solution to gauged $\Neql8$ supergravity can be obtained by using an uplift formula.  One writes the full metric on the warped product manifold, $M_{1,10}=M_{1,3}\times M_{7}$, as
\begin{equation}
\label{metrel}
{1\over\kappa^2} ds_{11}^2\eql \Delta^{-1} ds_{1,3}^2+a^2\,ds_7^2\,,
\end{equation}
where $M_7$ is a deformed  sphere, $S^7$, of unit radius with the metric, $ds_7^2=g_{mn}d\rho^m d\rho^n$,  and $a$ is the radius of the internal manifold. It will turn  out convenient below to have the   overall scale, $\kappa$, of the metric in eleven dimensions as an explicit parameter. 

The  inverse of the internal metric and the warp factor, $\Delta$, in \eqref{metrel} are given by the consistent truncation Ansatz  \cite{deWit:1984nz},
\begin{equation}
\label{inmetr}
\Delta^{-1}g^{mn}\eql   K_{AB}{}^m \,(\Gamma_{AB})^{IJ}\left(u_{ij}{}^{IJ}+v_{ijIJ}\right)
\left(u^{ij}{}_{KL}+v^{ijKL}\right)(\Gamma_{CD})^{KL}K_{CD}{}^n\,,
\end{equation}
 in terms of the scalar vielbein of $\Neql8$ supergravity and the Killing vectors, $K_{AB}=x^A\partial_B-x^B\partial_A$, on $S^7$, where $x^A$ are the Cartesian coordinates on $\RR^8$. The warp factor is defined by 
\begin{equation}\label{warpfactor}
\Delta^{-1}= \sqrt{\det(g^{mp}g^{\circ}_{pn}) }\,,
\end{equation}
where the metric $g^{\circ}_{mn}$ is that of the round sphere. The calculation of the actual deformed metric on $S^7$ using \eqref{inmetr} can be algebraically quite involved  (see, for example, \cite{Ahn:2002eh,{Corrado:2001nv}}), and in the following we will skip much of those technical details.

It is important to note that  the Ansatz \eqref{inmetr} includes the triality rotation on the scalar fields of $\cN=8$ supergravity from the $\rm SU(8) $ to the $\rm SL(8,\RR)$ basis \cite{Cremmer:1979up} that is implemented by the 
$\rm SO(8)$ gamma matrices,\footnote{We use the same convention as in \cite{Corrado:2001nv}.} $\Gamma_{AB}$. This  rotation must be taken into account when identifying the scalar fields in Section  \ref{scalarsection} with  the metric and flux deformations on $S^7$.  

The identification  is clear in the neighborhood of the $\rm SO(8)$-invariant fixed point corresponding to the $AdS_4\times S^7$ Freund-Rubin solution in eleven dimensions. Under the triality transformation, the $\bf 35_+$   scalar fields are mapped onto traceless symmetric tensors of  $\rm SO(8)$ and hence become ``metric deformations.'' Similarly, the $\bf 35_-$ pseudo-scalars map onto real self-dual tensors and are  identified as ``flux deformations.''
 
Finally, to make a connection with the discussion in Sections \ref{linearmodes}, we note that the  complex structure on the $\RR^8$ above that is preserved by $\rm SU(4)^-$ is
\begin{equation}\label{newcomplstr}
J_{AB}\eql {1\over 4}\,(\Gamma_{AB})^{IJ}J^{-}_{IJ}\,,
\end{equation}
where $J^-_{IJ}$ is the complex structure  in \eqref{complstr}. This gives an explicit identification of $\RR^8$, with the complex structure $J$ in \eqref{newcomplstr}, with the $\CC^4$ of the  harmonics in Section \ref{linearmodes}.

\subsubsection{The SU(4)$^-$ sector in eleven dimensions}
\label{truncsect}

As an illustration of the explicit uplift  formula \eqref{inmetr} for the metric  we consider    the $\rm SU(4)^-$ invariant background in Section \ref{su4trunc}, with the gauge field $B$ set to zero. We find that the $M$-theory metric has the warp factor
\begin{equation}\label{warpex}
\Delta\eql {1-\eta^2\over 1+\eta^2}(1-|\zeta|^2)^{2/3}\,,
\end{equation}
while the internal metric, written in  $\RR^8$, is given by 
\begin{equation}\label{themetric}
  ds_7^2\eql   (1-|\zeta|^2)^{1/3}\Big[\, dx^Adx^A-(x^AJ_{AB}\,dx^B)^2\,\Big]+{(1-\eta^2)^2\over (1-|\zeta|^2)^{2/3}(1+\eta^2)^2}\,(x^AJ_{AB}\,dx^B)^2\,.
\end{equation}
It is straightforward to verify using \eqref{hopfcoor} that the term in the square bracket is the metric on $\CC\PP^3$, while 
$x^AJ_{AB}dx^B=d\psi+A$. Note that as required by  the $\rm SU(4)$ invariance, the metric preserves the Hopf fibration, but scales the $\CC\PP^3$ base relative to the Hopf fiber. This class of metrics has been introduced in \cite{Pope:1984jj}  and the metric in Section \ref{popewarnersol} is a particular example. The metric derived   here is in agreement with the beautiful result for the uplift of the entire sector, including the vector field, obtained recently in \cite{{Gauntlett:2009zw},Gauntlett:2009bh}.

Since there is no comparably practical formula for the   flux in the uplift of $\cN=8$ supergravity to M-theory,%
\footnote{Implicit formulae for the flux have been derived in \cite{de Wit:1986iy}, but those are rather difficult to use.}
we conclude this section by   quoting the result for the flux in \cite{{Gauntlett:2009zw},Gauntlett:2009bh} in the subsector discussed in  Section \ref{sufourcrit}, that is with  $B=0$,  $\eta=0$, and  with  $\zeta$ real.  We find 
\begin{equation}\label{thefluxx}
\begin{split}
{1\over \kappa^3} \,F_{(4)}\eql    {1\over 2a}\,{(6-8\,\zeta^2)\over (1-\zeta^2)^2}\, {\rm vol}_{M_{1,3}}+ {a^3\over 4} \, d(\zeta\,\Omega+{\rm c.c.})\,,
\end{split}
\end{equation}
where $\Omega$ is defined in \eqref{theomegaz}.  

Note that by setting 
\begin{equation}\label{getpwpt}
\zeta\eql {1\over \sqrt2}\,,\qquad a\eql {4\over \sqrt 3}\,L\,,\qquad \kappa\eql {1\over 2^{1/3}}\,,
\end{equation}
we obtain the PW solution in \eqref{popewarmet} and \eqref{fluxpw} as an uplift of the critical point in the $\rm SU(4)^{-}$ invariant sector in Section \ref{su4trunc} with $a^2=2/g^2$.

\subsubsection{The unstable metric modes}
\label{metricmodes}

An agreement between the result for the metric \eqref{themetric} obtained using the uplift fromula \eqref{inmetr} and the metric in the uplift of the entire $\rm SU(4)^-$ invariant sector in \cite{{Gauntlett:2009zw},Gauntlett:2009bh} provides yet another confirmation of the consistent truncation Ansatz in \cite{deWit:1984nz}. Since the unstable  modes arise from the scalar fields of $\cN=8$ supergravity, we will now use \eqref{inmetr} to determine the exact metric fluctuations corresponding to those modes.  

From the perspective of the metric Ansatz \eqref{inmetr}, the non-vanishing  metric deformations due to the flux modes \eqref{metrhar} are not that surprising. At the $\rm SO(8)$ point, the scalar vielbeins $u_{ij}{}^{IJ}$, $u^{ij}{}_{KL}$, $v_{ijIJ}$ and $v^{ijKL}$ in 
\eqref{inmetr}
correspond to the group element $\widetilde \cV$ defined in \eqref{quadexpan} with the $G_i$'s given by the subset of generators $\Sigma_{IJKL}$ along the $\bf 35_-$ flux components. Those generators in  \eqref{inmetr} are projected directly onto the triality rotated Killing vectors, $K_{IJ}=K_{AB}(\Gamma _{AB})^{IJ}$. This results in terms   of the form $K_{IJ}\Sigma_{IJKL}K_{KL}$ that yield a vanishing correction to the metric at the linear order. On the other hand, away from the origin of the coset, the scalar vielbeins are determined by the group element  $\cV=\widetilde\cV{\cV}_0 $ in \eqref{expandvar} and the projection of the flux components onto the Killing vectors in \eqref{inmetr} is ``rotated'' by the vielbeins  corresponding to $\cV_0$. One could argue using the explicit form of $\cV_0$ at the $\rm SU(4)^-$ critical point that  the correction to the metric must be of the form \eqref{metrhar}.

This expectation is confirmed by an explicit calculation of the metric fluctuations corresponding to the unstable scalar modes, $\widetilde \cV$, in the $\bf 20'$ of $\rm SU(4)^-$, for an arbitrary value  of $\zeta$ that parametrizes  the scalar manifold of the truncated sector in Sections \ref{sufourcrit} and  \ref{truncsect}. Specifically, we use \eqref{inmetr} to evaluate the metric corresponding to the point  $\widetilde\cV{\cV}(\zeta) $ and then expand it to the first order in scalar fluctuations. To obtain the actual fluctuations of the metric we also  calculate the warp factor and then invert the metric, with   the last step  being straightforward given the exact  unperturbed metric  in  \eqref{themetric}.

We find that for the scalar modes in $\bf 20'$, the warp factor is not corrected at the linear order and that the Hopf fibration structure in  the metric   \eqref{themetric} is preserved. The only deformation  of the metric  at the linear order is for the metric components along  $\CC\PP^3$ and is of the form
 \begin{equation}\label{genmetrflct}
\delta g_{z_iz_j}\eql \zeta (1 - \zeta^2)^{1/3} {e^{4i\psi}\over (1-\sum_k |z_k|^2)^2}\,\delta h_{z_iz_j}\,,
\end{equation}
and similarly for the antiholomorphic components, $\delta g_{\bar z_i\bar z_j}$. In the coordinates that we are using, $\delta h_{z_iz_j}$ are holomorphic, and in fact are  given by second order polynomials in the $z_i$'s. Hence it is quite straightforward to recast \eqref{genmetrflct} in terms of harmonics and recover the same result as in \eqref{metrhar} that was obtained by solving the equations of motion. 

The  dependence on $\zeta$ in \eqref{genmetrflct} shows  that for $\zeta=0$, that is at  the $\rm SO(8)$ critical point, there is no linear correction to the metric. This is consistent with the interpretation of these modes as ``flux'' modes that mix with the metric only away from the $\rm SO(8)$ point.

\subsection{Comments and generalizations}

An obvious question is whether the  M-theory solution based on the PW deformation of $S^7$ can be stabilized by either judiciously projecting out the troublesome modes or, more generally, by replacing the sphere with another Sasaki-Einstein manifold, $\rm SE_7$, with a  K\"ahler-Einstein base, $\rm KE_6$, other than  $\CC\PP^3$, or by some combination of both.

In the first case, an obvious mechanism to try is an orbifold construction based on a discrete  subroup of $\rm SU(4)\subset SO(8)$. It is easy to argue that this cannot  work if the orbifold group is abelian. Indeed, such an abelian subgroup can be conjugated to the maximal torus of $\rm SU(4)$ and there are  always at least two  unstable modes in $\bf 20'$ that are neutral and hence  will not be projected out. To see this explicitly,  recall that  $\bf 20'$ can also be  realized  as symmetric, real, traceless matrices with the obvious action of $\rm SO(6)\simeq SU(4)$. The maximal torus is then the block diagonal $\rm SO(2)^3\subset SO(6)$ and the neutral modes, that are manifestly $\rm SO(2)^3$ invariant, are given by the diagonal matrices ${\rm diag} (a,a,b,b,c,c)$ with $c=-(a+b)$.  

A more difficult option is to consider  non-abelian orbifolds, but for that one would have to choose some exotic subgroup of $\rm SU(4)$ as  the simplest choice of using the Weyl group does not work given that ${\bf 20'}$   has Weyl singlets. It is also important to keep in mind the interesting work \cite{Dymarsky:2005nc}, where it was shown that non-supersymmetric twisted sectors generically destabilize a solution.  Finally, there are arguments in \cite{Horowitz:2007pr} that non-supersymmetric orbifolds of $AdS_p\times S^q$ solutions have non-perturbative instabilities, and it would be rather surprising if the same would not hold for  more general solutions with background fluxes. 

The starting point of the more general construction of Pope-Warner type solutions \cite{Pope:1984jj,Pope:1984bd} is to take the canonical $\rm U(1)$ fibration over a general K\"ahler-Einstein six-manifold, $\rm KE_6$.  As  noted in \cite{Gauntlett:2009dn, Gauntlett:2009bh}, if one canonically normalizes the $\rm U(1)$ fiber relative to the $\rm KE_6$ base,  the resulting background is  a Sasaki-Einstein seven manifold,%
\footnote{For a recent comprehensive review of Sasaki-Einstein manifolds in the context of string theory see \cite{Sparks:2010sn}.}
 $\rm SE_7$, and hence the cone over it   is a Calabi-Yau four-fold, $\rm CY_4$.   The solution involves taking the holomorphic $(4,0)$ form on $\rm CY_4$, descending it to   $\rm KE_6$ and using it as the internal  flux.  In the presence of the flux, the metric on  $\rm SE_7$ gets deformed  by streching the $\rm U(1)$ fiber relative to the $\rm KE_6$ base, which remains the same.

The general pattern that emerges from the results in this section suggests that the source of  instabilities might be primitive harmonic $(2,2)$ forms on the $\rm CY_4$, which, through their coupling to the background flux, will source metric variations that deform  the complex structure on the $\rm KE_6$.  The equations of motion  are then being solved with flux sources that involve the background flux interacting with linearized $(2,2)$, $(3,1)$ and $(1,3)$ forms and it is this interaction that sources the deformation of the metric and the violation of the underlying hermitian structure. We therefore suspect that if there are primitive harmonic $(2,2)$ forms   on the $\rm CY_4$, then these could lead to problematic instabilities.  

Our analysis of the instability  also leads to two other important conclusions.  First, one cannot look inside $\Neql 2$ truncated theories alone and, secondly, one cannot restrict to complex manifolds at the non-trivial fixed point.   The instabilities of the $\rm SU(4)^-$ point illustrate this observation very clearly.  It remains to be seen whether this instability is a general feature of the Pope-Warner construction or whether the instability is evitable by a suitable choice of the $\rm KE_6$ base and perhaps some more sophisticated orbifold. Our analysis indicates very clearly where one might expect to find dangerous terms.

\section{Conclusions}
\label{sec:conclude}

Motivated by recent work on holographic superconductors, we have computed the scalar mass spectrum in $\cN=8$ gauged supergravity of all critical points in the $\rm SU(3)$-invariant sector. The $\rm SO(8)$, $\rm G_2$ and $\rm SU(3)\times U(1)$-invariant critical points are all supersymmetric and stable, while the $\rm{SO(7)^+}$, $\rm SO(7)^-$ and $\rm SU(4)^-$ solutions are all non-supersymmetric and unstable due to certain scalars that violate the BF bound. Particular attention has been paid to the $\rm SU(4)^-$ solution which has recently been proposed as a  holographic dual to the  ground state of a CFT which exhibits a superconducting phase at finite temperature. Our results suggest that due to the instability in the supergravity solution, there is in fact no stable holographic dual CFT.

The PW solution, that uplifts  the $\rm SU(4)^-$ critical point to M-theory, has an interesting generalization in which $S^7$ is replaced with a more general $\rm SE_7$ manifold.  The Lagrangian (\ref{sufouraction}) remains a consistent truncation of this theory \cite{Gauntlett:2009zw}, but there is most definitely a  rich  spectrum of modes outside the truncation. It is reasonable to conjecture that the unstable modes  we have identified will have analogues  in this more general class of backgrounds, but an eleven-dimensional computation of the spectrum would be necessary for this to be rigorously settled. More precisely, we conjecture that some  primitive, harmonic $(2,2)$ forms on the Calabi-Yau fourfold, which is a cone over $\rm SE_7$, might give rise to four-dimensional scalars which violate the BF bound.

While we have focused in this work on one very specific example of a non-supersymmetric solution to eleven-dimensional supergravity, our results are symptomatic of a much wider problem in string theory. Without supersymmetry to protect the vacuum state, generically, non-supersymmetric solutions are unstable. It is not sufficient to demonstrate stability within a certain truncated sector where one may find a putative vacuum since embedding such a vacuum into string theory typically produces a rich spectrum of low lying modes. This issue is of some pertinence with regards to holographic attempts at studying condensed matter systems where typically one explicitly breaks supersymmetry. This should be contrasted with another aspect of applied AdS/CFT, namely the use of the gravity dual of $\cN=4$ SYM to study heavy-ion physics  (for a review and references see \cite{Gubser:2009md, Gubser:2009fc}) where the $\cN=4$ theory clearly has a stable ground state and there is a well established holographic dictionary. 

We have also provided a detailed description of the $\rm SU(3)$-invariant sector of $\cN=8$ gauged supergravity as a four dimensional, $\cN=2$, $\rm U(1)\times U(1)$ gauged supergravity with one vector multiplet and one hypermultiplet, following the formal structures summarized in \cite{Andrianopoli:1996vr,Andrianopoli:1996cm}. It is interesting that the Lagrangian of the $\rm SU(4)$-invariant sector, (\ref{sufouraction}), can be embedded into such a supergravity theory in two different ways. In addition to the description provided here, a different $\cN=2$ theory was constructed in \cite{Gauntlett:2009zw} by keeping the $\rm SU(4)$-invariant modes from KK towers other than those which give rise to gauged supergravity, in particular they keep the so-called {\it breathing mode}. The difference between the two constructions lies only in the particular gaugings and this subsequently gives rise to different interactions via the scalar potential (\ref{frepot}). Interesting related work on many of these issues for the IIB version of this story has appeared recently in \cite{Liu:2010ya}.

\bigskip
\bigskip
\bigskip
\leftline{\bf Acknowledgements}
\smallskip
We would like to thank  Davide Forcella, Jerome Gauntlett and Stanislav Kuperstein for useful conversations. This work was supported in part by DOE grant DE-FG03-84ER-40168. The work of N.H has been supported in part by the DSM CEA-Saclay, by the ANR grants BLAN 06-3-137168 and JCJC ERCS07-12 and by the Marie Curie IRG 046430.


\vfill\eject


\begin{appendices}

\appendix
\section{The other critical points}

\renewcommand{\theequation}{A.\arabic{equation}}
\renewcommand{\thetable}{A.\arabic{table}}
\setcounter{equation}{0}
\label{appendixA}

\begin{table}[h]
\begin{center}
\begin{tabular}{@{\extracolsep{6 pt}}c   c c c  c }
\toprule
\noalign{\smallskip}
$\rm G$  \hspace{-10 pt} &   $z$ & $\zeta_1$  & $\zeta_2$ &  $\cP_*/g^2$  \\
\noalign{\smallskip}
\midrule
\noalign{\medskip}
SO(8)  & 0 & 0 & 0 &  $-6 $   \\
\addlinespace[10 pt]
$\rm SO(7)^+$ & $\displaystyle -{1-5^{1/4}\over 1+5^{1/4}}$ &$\displaystyle{1\over 2}\big(3-\sqrt 5\big)$ & 0    &  $-2\cdot 5^{3/4}\approx -6.687$ \\
\addlinespace[10 pt]
$\rm SO(7)^-$  & $-i\,(2-\sqrt 5)$ &  0 &   $\displaystyle {1\over\sqrt 5}$ &  $\displaystyle-{25\over 8}\sqrt 5\approx -6.988$ \\
\addlinespace[10 pt]
$\rm G_2$  & $z_{(2)}$ & $2-\sqrt 3$  & $\displaystyle {1\over \sqrt{3+2\sqrt 3}}$ \hspace{- 10pt}&   $\displaystyle  -{216\over 25}\sqrt{{2\over 5}\sqrt 3}\approx -7.192$   \\
\addlinespace[10 pt]
$\rm SU(3)\times U(1)$ &  $2-\sqrt 3$   & 0 & $\displaystyle {1\over \sqrt{3}}$ &    $\displaystyle -{9\over 2}\sqrt 3\approx -7.794$  \\
\addlinespace[10 pt]
$\rm SU(4)^-$  & 0 &  0 & $\displaystyle{1\over \sqrt{2}}$  &  $-8 $ \\
\addlinespace[10 pt]
\midrule
\addlinespace[10 pt]
& \multicolumn{4}{l}{$\displaystyle z_{(2)}\eql {1\over 4}\left(1-{i\over 3^{1/4}}\,\sqrt{2+\sqrt 3}\,\right)\big(3+\sqrt 3-3^{1/4}\sqrt{10}\,\big)$} \\
\addlinespace[10 pt]
\bottomrule
\end{tabular}
\caption{All critical points in the $\rm SU(3)$-invariant sector.  }
\label{criticaltable}
\end{center}
\end{table}

We have determined the masses of all  scalar fluctuations around all of the $\rm SU(3)$-invariant critical points using the method described in Section \ref{QuadFlucts}.   In the cases with higher symmetry, or with supersymmetry, our calculation reproduces   known  results  in the literature for the $\rm SO(8)$ \cite{Casher:1984ym,Sezgin:1983ik} (see also \cite{Duff:1986hr} for a review), $\rm SO(7)^\pm$ \cite{deWit:1983gs}  and  $\rm SU(3)\times U(1)$  \cite{Nicolai:1985hs,Klebanov:2008vq,Klebanov:2009kp} points. The results for the supersymmetric $\rm G_2$ point and the non-supersymmetric $\rm SU(4)^-$ point are  new.

\begin{table}[t]
\begin{center}
\begin{tabular}{@{\extracolsep{25 pt}}r c c c  }
\toprule
\noalign{\smallskip}
$n_M$ \hspace{-10 pt} & $M^2L^2$ & $\rm SO(8)$ irreps & $\rm SU(3)$ singlets  \\
\noalign{\smallskip}
\midrule
\noalign{\medskip}
35 & $-2 $ & $\bf 35_+$ & 3    \\
\noalign{\medskip}
35 & $-2 $ & $\bf 35_- $ & 3    \\
\noalign{\smallskip}
\bottomrule
\end{tabular}
\caption{The $\rm SO(8)$ point.  }
\label{tblso(8)}
\end{center}
\end{table}

\begin{table}[t]
\begin{center}
\begin{tabular}{@{\extracolsep{15 pt}}r c c c c}
\toprule
\noalign{\smallskip}
$n_M$ \hspace{-10 pt} & $M^2L^2$ & $\rm SO(7)^{\pm}$ irreps & $\rm SU(3)$ singlets & Stability\\
\noalign{\smallskip}
\midrule
\noalign{\medskip}
1 & \phm $ 6 $ & $\bf 1 $ & 1   & S \\
\noalign{\medskip}
7 & \phm $0 $ & $\bf 7 $ & 1  & S  \\
\noalign{\medskip}
35 & $\displaystyle -{6\over 5} $ & $\bf 35_\mp $ & 3  & S \\
\noalign{\medskip}
27 & $ \displaystyle -{12\over 5} $ & {\bf 27} & 1 & U \\
\noalign{\smallskip}
\bottomrule
\end{tabular}
\caption{The $\rm SO(7)^{\pm}$ points.  }
\label{tblsoseven}
\end{center}
\end{table}

The masses of scalar modes are defined by the canonical form of the scalar field equation in $AdS_4$,
\begin{equation}\label{scaleqs}
\left(\Box_{AdS}-M^2 \right)\phi\eql 0\,.
\end{equation}
For the canonically normalized kinetic term, the value of the mass is related to the eigenvalue of the second variation of the potential,
\begin{equation*}
M^2L^2\eql -3 \,\left({\cP''(\phi_0)\over \cP(\phi_0)}\right)\,,
\end{equation*}
where $L$ is the radius of the $AdS_4$ critical point of interest. In terms of the dimensionless mass parameter, $M^2L^2$,  the stability condition \eqref{Bfcrit} in $AdS_4$ reads
\begin{equation}\label{sstability}
M^2L^2~\geq~ -{9\over 4}\,.
\end{equation}
With this normalization the conformal dimension, $\Delta$, of the operator in the holographic CFT dual to a scalar of mass $M^2$ is given by a solution to the equation
\begin{equation}
 M^2L^2\eql \Delta (\Delta-3)\,.
 \label{massreln}
\end{equation}

We list all critical points in the $\rm SU(3)$-invariant sector by their symmetry group $\rm G$ in Table~\ref{criticaltable}. The table also gives the  values of the  background fields, $z$, $\zeta_1$ and $\zeta_2$  and of the cosmological constant, $\cP_*=\cP(\phi_0)$, at each critical point. Note that we have picked a gauge in which the charged scalars $\zeta_1$ and $\zeta_2$  are real.  The reader should consult the original table in \cite{Warner:1983vz} for more details.

\begin{table}[t]
\begin{center}
\begin{tabular}{@{\extracolsep{25 pt}}r c c c }
\toprule
\noalign{\smallskip}
$n_M$ \hspace{-10 pt} & $M^2L^2$ & $\rm G_2$ irreps & $\rm SU(3)$ singlets\\
\noalign{\smallskip}
\midrule
\noalign{\medskip}
1 & $(4+\sqrt 6\,)  $ & $\bf 1 $ & 1    \\
\noalign{\medskip}
1 & $(4-\sqrt 6\,)
 $ & $\bf 1 $ & 1    \\
\noalign{\medskip}
14 & $0 $ & $\bf 7\oplus 7 $ & 1+1    \\
\noalign{\medskip}
27 & $ -{1\over 6}(11-\sqrt 6\,)$ & {\bf 27} & 1 \\
\noalign{\medskip}
27 & $ -{1\over 6}(11+\sqrt 6\,)$ & {\bf 27} & 1 \\
\noalign{\smallskip}
\bottomrule
\end{tabular}
\caption{The $\rm G_2$ point.  }
\label{tblgtwo}
\end{center}
\end{table}

\begin{table}[t!]
\begin{center}
\begin{tabular}{@{\extracolsep{25 pt}}c c c c c }
\toprule
\noalign{\smallskip}
$n_M$ & $M^2L^2$ & $\rm SU(4)^-$ irreps & $\rm SU(3)$ singlets & Stability\\
\noalign{\smallskip}
\midrule
\noalign{\smallskip}
2 & \phm 6 & $\bf 1\oplus 1$ & 2 &  S\\
\noalign{\medskip}
&  & $\bf 15 $ & 1 &  \\
28   & \phm 0  & $\bf 6\oplus\overline 6$ & & S\\
   &   & $\bf 1$ & 1 & \\
\noalign{\smallskip}
20 & $\displaystyle -{3\over 4}$ & $\bf 10\oplus \overline{10}$ & 1+1 & S \\   
\noalign{\medskip}
20 & $-3$ &  $\bf 20'$ & & U \\
\bottomrule
\end{tabular}
\caption{The $\rm SU(4)^-$ point.}
\label{tblsufour}
\end{center}
\end{table}

\begin{table}[t!]
\begin{center}
\begin{tabular}{@{\extracolsep{16 pt}}c c c c   }
\toprule
\noalign{\smallskip}
$n_M$ & $M^2L^2$ & $\rm SU(3)\times U(1)$ irreps & $\rm SU(3)$ singlets\\
\noalign{\smallskip}
\midrule
\noalign{\smallskip}
1 & $3+\sqrt{17}$ & ${\bf 1}(0)$ & 1 \\
\noalign{\medskip}
3 & 2 & $\bone(+1)\oplus \bone(0)\oplus\bone(-1)$  & $1+1+1$\\
\noalign{\medskip}
&  & $\bthr(+{2\over 3})\oplus \overline\bthr(-{2\over 3})$ & \\
\noalign{\smallskip}
19 & 0  &  $\bthr(-{1\over 3})\oplus\bthr(-{1\over 3})\oplus \overline\bthr(+{1\over 3})\oplus\overline\bthr(+{1\over 3})$ & \\
\noalign{\smallskip}
 & & $\bone(0)$ & 1 \\
\noalign{\medskip}
1 & $3-\sqrt{17} $ & $\bone(0)$ & 1 \\
\noalign{\medskip}
18 & $\displaystyle -{14\over 9}$ & ${\bf 6}(+{1\over 3})\oplus\bthr(-{1\over 3})\oplus\overline\bthr(+{1\over 3})\oplus  \overline {\bf 6}(-{1\over 3})$ \\
\noalign{\medskip}
16 & $-2$ & ${\bf 8}(0)+{\bf 8}(0)$ \\
\noalign{\medskip}
12 & $\displaystyle -{20\over 9}$ & ${\bf 6}(-{2\over 3})\oplus \overline{\bf 6}(+{2\over 3}) $ \\
\noalign{\medskip}
\bottomrule
\end{tabular}
\caption{The $\rm SU(3)\times U(1) $ point.}
\label{tblsuthree}
\end{center}
\end{table}

The results of our calculation of the masses are summarized in Tables \ref{tblso(8)} -\ref{tblsuthree}. The first column in each table gives the number of scalar modes, $n_M$, for a given value of $M^2L^{2}$ that is listed  in the second column. In the third column we give the corresponding representations of the symmetry group, $\rm G\subset SO(8)$, at the critical point. The fourth column gives the number of modes that are $\rm SU(3)$ singlets. We have verified that their masses agree with the ones calculated by expanding  the truncated action of the $\rm SU(3)$-invariant sector in Section \ref{suthreesector}. When there are unstable modes present,  we indicate whether a mode  is stable (S) or unstable (U) in the last column.

As expected, all supersymmetric points: $\rm SO(8)$, $\rm G_2$ and $\rm SU(3)\times U(1)$ are perturbatively stable, while all non-supersymmetric points: $\rm SO(7)^\pm$ and $\rm SU(4)^-$ are unstable. The massless scalars at all critical points, except $\rm SU(4)^{-}$, correspond precisely to the Goldstone bosons due to the spontaneous breaking of $\rm SO(8)$ to the corresponding subgroup at the critical point. At the $\rm SU(4)^{-}$ invariant critical point there are 13 Goldstone bosons in the $\bf 6\oplus\overline{6} \oplus 1$ of $\rm SU(4)^-$  and 15 extra zero modes in the $\bf 15$ of $\rm SU(4)^-$.

The decomposition of scalar modes with a given mass  into irreducible representations of $\rm G$ for most critical points is unambigous  just by matching the degeneracies with  dimensions of  representations of $\rm G$ that arise in the branching\footnote{For convenience, we have summarized the relevant branching rules in Appendix \ref{appendixC}.} of $\bf 35_+$ and $\bf 35_-$ of $\rm SO(8)$ under $\rm G$. However for $\rm SU(4)^-$ this approach has to be supplemented by the branching with respect to $\rm SU(3)$ and  the direct calculation of masses in the $\rm SU(3)$-invariant sector. This works for all but some modes  at the  $\rm SU(3)\times U(1)$ point, which we will   discuss now  in more detail.

The $\rm SU(3)\times U(1)$-invariant solution, and its representation theory, was studied thoroughly in  \cite{Nicolai:1985hs}, where the decomposition of all modes into supermultiplets of $\rm SU(3)\times OSp(2,4)$  was given (see also \cite{Klebanov:2008vq,Klebanov:2009kp} for a more recent discussion). The scalar masses can be read off from the superconformal dimensions and  $\cR$ charges of those supermultiplets as follows:

For short multiplets, the masses are given by the familiar formula from holography \eqref{massreln} and $\Delta$ is related to the $\rm U(1)$ $\cR$-charges by $\Delta = 2 |\cR|$ with the $\rm U(1)$ charge normalizations employed in Table \ref{tblsuthree}.  This explains the masses for the $18$ and $12$ scalar modes in massive hypermultiplets.   
In a long massive vector multiplet there are five scalars  \cite{Nicolai:1985hs} and they have $\Delta = \Delta_0$, $\Delta = \Delta_0+1$ and $\Delta = \Delta_0+2$, with degeneracies $1,3$ and $1$ respectively.    The exact value of $\Delta_0$ is  $\Delta_0 = \frac{1}{2} (1 + \sqrt{17})$.  This explains the first five singlet entries in the table, which belong to massive vector multiplets.  The   sixteen modes which  have the $M^2L^2$  value of exactly  $-2$ are the ``massless" modes in the  $\Neql2$, $\rm SU(3)$ vector multiplet and so lie in the $ {\bf 8}(0) \oplus {\bf 8}(0)$ and  are conformally coupled. Finally, the 19 Goldstone modes necessarily have no mass terms.

\section{The $\rm\bf SU(3)$-invariant sector in canonical $\boldmath \Neql2$ form}
\renewcommand{\theequation}{B.\arabic{equation}}
\setcounter{equation}{0} 
\label{appendixB}

For completeness we now re-write the bosonic action of the $\Neql2$ truncation derived in Section~\ref{suthreesector} in terms of the canonical form of $\cN=2$ supergravity  as given in \cite{Andrianopoli:1996vr}.%
\footnote{For a comprehensive review of the vast subject of  matter coupled $\cN=2$ supergravity, we refer the reader to \cite{Andrianopoli:1996vr,Andrianopoli:1996cm} and the references therein, and to     \cite{Ceresole:1995ca} and \cite{deWit:1995jd} for a succinct summary of the main features of these theories. The $\rm SU(3)$-invariant truncation of  ungauged  $\cN=8$ supergravity in four dimensions was discussed in \cite{Cecotti:1988qn}, where it was shown that the truncation could be recast in the canonical form.
}

The main result of \cite{Andrianopoli:1996vr} is to show how the complete action of $\cN=2$ supergravity is determined in terms of the following geometric quantities on the special K\"ahler and quaternionic K\"ahler scalar manifolds:
\bigskip

\noindent
\begin{itemize}
\item [I.]  Special K\"ahler
\begin{equation}\label{SKman}
{\cal M}_{\rm SK}\eql {\rm SU(1,1)\over U(1)\times U(1)}\,.
\end{equation}
\begin{itemize}
\item [(i)] holomorphic sections: $X^\alpha$, $\alpha=0,1$,
\item [(ii)] holomorphic prepotential, $F(X^0,X^1)$.
\end{itemize}

\item[II.] Quaternionic K\"ahler
\begin{equation}\label{Qman}
{\cal M}_{\rm QK}\eql {\rm SU(2,1)\over SU(2)\times U(1)}\,.
\end{equation}
\begin{itemize}
\item [(i)] metric, $g_{ab}$,
\item [(ii)] quaterninionic prepotentials, $\cP_\alpha^i$, $i=1,2,3$, for the Killing vectors $K_\alpha$.
\end{itemize}
\end{itemize}

The initial problem is to identify correctly the truncated fields from \cite{de Wit:1982ig} with the corresponding fields in \cite{Andrianopoli:1996vr}. To do this properly, one needs to understand all conventions in   \cite{de Wit:1982ig} and \cite{Andrianopoli:1996vr}, in particular those for the fermion fields, and then truncate not just the action, but also the supersymmetry variations. The latter are needed to properly disentagle fields in different multiplets. We will present the details of these calculations below. 

\subsection{The special K\"ahler manifold}

Start with the fact that we can identify  the complex structure and the K\"ahler metric on ${\cal M}_{\rm SK}$ in \eqref{SKman} by comparing the kinetic terms  \eqref{kinetic} with the standard form of the  $\cN=2$ action:\footnote{Note that the authors of \cite{Andrianopoli:1996vr} use the opposite convention for the metric signature and have a different normalization of the gauge fields.}
\begin{equation}
e^{-1}\cL\eql - g_{z\bar z}\, \nabla_\mu z  \nabla^\mu \bar z \,.
\end{equation}
This yields
\label{a}
\begin{equation}
\label{Kmetr}
g_{\zet \bzet} \eql   { 3\over (1-|z|^2)^2}\,,
\end{equation}
and identifies $\zet$ as a  holomorphic coordinate and the corresponding K\"ahler potential as:
\begin{equation}
\label{Kahlpot}
K(\zet,\bzet)\eql -3 \log(1-|\zet|^2)+f(\zet)+\bar f(\bzet)\,.
\end{equation}
where $f(z)$ is an, as yet, arbitrary analytic function.

The holomorphic sections, $X^\alpha(z)$, $\alpha=0,1$, can be read off from the supersymmetry transformations of the vector fields given in (4.23) of \cite{Andrianopoli:1996vr}:
\begin{equation}
\label{varAfre}
\delta A_\mu^\alpha\eql 2 L^\alpha\overline\psi_\mu^A\epsilon^B\epsilon_{AB}+f^\alpha _z\overline \lambda^A\gamma_\mu\epsilon^B\epsilon_{AB}+{\rm h.c.}\,,
\end{equation}
where
\begin{equation}
\label{Lsect}
L^\alpha(\zet,\bzet)\eql e^{K(\zet,\bzet)/2}X^\alpha(\zet) \,,
\end{equation}
and 
\begin{equation}
\label{fvielb}
f^\alpha_z(\zet,\bzet)\eql \left(\partial_z+{1\over 2}\partial_z K\right)L^\alpha\,.
\end{equation}
This must  be compared with the truncation of 
\begin{equation}
\label{varvecdwn}
\delta A_\mu^{IJ}\eql -(u_{ij}{}^{IJ}+v_{ijIJ})\left(\overline \varepsilon_k\gamma_\mu\chi^{ijk}+2\sqrt 2\, \overline \varepsilon^i\psi_\mu^i\right)+{\rm h.c.}\,,
\end{equation}
in (3.3) of \cite{de Wit:1982ig}. 

From the embedding of $\rm SU(3) \hookrightarrow SO(8) \hookrightarrow SU(8)$, we find that the $\rm SU(3)$-invariant gravitino components are  $\psi_\mu^7$ and $\psi_\mu^8$, and similarly the two parameters of the supersymmetry are $\varepsilon^7$ and $\varepsilon^8$.   The truncation of the gravitino contribution in \eqref{varvecdwn} yields  (up to an overall sign)
\begin{equation}\label{varwpsi}
\begin{split}
\delta A_\mu^{12}&\eql \sqrt{2}{\zet(1+\zet)\over (1-|\zet|^2)^{3/2}} (\bar\varepsilon^8\psi_\mu^7-\bar\varepsilon^7\psi_\mu^8)\,,\\[6 pt] 
\delta A_\mu^{78} & \eql \sqrt{2} {(1+\zet^3)\over (1-|\zet|^2)^{3/2}}(\bar\varepsilon^8\psi_\mu^7-\bar\varepsilon^7\psi_\mu^8)\,.
\end{split}
\end{equation}
By setting $\zet=0$, we identify $A_\mu^{78}$ as the graviphoton and $A_\mu^{12}=A_\mu^{34}=A_\mu^{56}$ as the vector of the vector multiplet.  
We see that \eqref{varwpsi} indeed has the correct  structure and we identify,
\begin{equation}
\label{holsectL}
L^0(\zet,\bzet)\eql c_L \sqrt{2}\,{(1+\zet^3)\over  (1-|\zet|^2)^{3/2}}\,,\qquad L^1(\zet,\bzet)\eql c_L\,\sqrt 6\,{\zet(1+\zet)\over  (1-|\zet|^2)^{3/2}}\,,
\end{equation}
from which we find the holomorphic sections
\begin{equation}
\label{holsect}
X^0(z)\eql c_X \sqrt{2}\,(1+\zet^3)\,,\qquad X^1(z)\eql c_X\,\sqrt 6\,\zet(1+\zet)\,,
\end{equation}
where $c_L$ and $c_X$ are are constants introduced to account for any signs and normalizations. 
Note that requiring that $X^\alpha(z)$ be holomorphic functions of $z$ fixes the K\"ahler gauge in \eqref{Kahlpot} to $f(z)=\log f_0$ so that
\begin{equation}\label{cons}
c_L\eql c_X |f_0|\,.
\end{equation}

The truncation of \eqref{varvecdwn} also allows identification of the spin-$1\over 2$ fields. By group theory, there are four  $\rm SU(3)$-invariant  spin-$1\over 2$ fields that can be parametrized by the components $\chi^{245}$, $\chi^{246}$, $\chi^{567}$ and $\chi^{568}$, respectively  \cite{Nicolai:1985hs}. The variation of the vector field reads:
\begin{equation}
\label{varchi}
\begin{split}
\delta A_\mu^{12}&\eql-{(1+2\zet+2\zet\bzet+\zet^2\bzet)\over (1-|\zet|^2)^{3/2}} (\bar\varepsilon^7\chi^{567}+\varepsilon^8\chi^{568})+{\rm h.c.}\,,\\[6 pt]
\delta A_\mu^{78}& \eql -3{(\bzet+\zet^2)\over (1-|\zet|^2)^{3/2}} (\bar\varepsilon^7\chi^{567}+\varepsilon^8\chi^{568})+{\rm h.c.}\,.
\end{split} 
\end{equation}
This identifies $\chi^{567}$ and $\chi^{568}$ as the fields in the vector multiplet. A direct comparison with \eqref{varAfre} is a bit trickier as one first needs to rotate to the complex basis for the spinors. 

Having identified the graviphoton and the gauge field in the vector multiplet, we can use the truncation of the Maxwell action to calculate the prepotential. All we need to do is to rewrite the action \eqref{maxwell} in the canonical form as in \cite{Andrianopoli:1996vr},
\begin{equation}
\label{maxwellsec6}
\begin{split}
e^{-1}\,\cL_{Max.} & \eql - \ds\frac{i}{4} \left( \cN_{\alpha\beta}  F_{\mu\nu}^{+ \alpha} F^{+\beta}{}^{\mu\nu}-\overline\cN_{\alpha\beta}  F_{\mu\nu}^{-\alpha} F^{-\beta}{}^{\mu\nu} \right)\\[6pt]
&\eql  \ds\frac{1}{4} \left( {\rm Im} \, \cN_{\alpha\beta} \,F_{\mu\nu}^\alpha F^{\beta\,\mu\nu}-i\,{\rm Re}\,\cN_{\alpha\beta} \,F_{\mu\nu}^\alpha\widetilde F^{\beta\,\mu\nu}\right)
\,.
\end{split}
\end{equation}
This gives: 
\begin{equation}
\label{Nmatr}
\begin{split}
 \cN_{00} & \eql  -c_N{i\over 4} {\left(1+2\bzet+3 \zet\bzet+3\zet^2+2\zet^3+\zet^3\bzet\right)\over (1+\zet)^2\left (1-2\zet+2\bzet-\zet \bzet\right)}\,,\\[6pt]
\cN_{11}& \eql 
-c_N{i\over 4} {\left(1-2\bzet+3 \zet\bzet+3\zet^2-2\zet^3+\zet^3\bzet\right)\over (1+\zet)^2\left (1-2\zet+2\bzet-\zet \bzet\right)}
\,,\\[6pt]
\cN_{01}&  \eql c_N{i\over 2}{\sqrt{3}\,\zet(1+\zet\bzet)\over (1+\zet)^2\left (1-2\zet+2\bzet-\zet \bzet\right)} \,,\\[12pt]
 \cN_{10} & \eql\cN_{01}\,,
\end{split}
\end{equation}
where $c_N$ is constant allowed by a rescaling of the vector fields.

If a prepotential, $F(X^0,X^1)$, exists, it must be a homogenous function of degree 2 of the holomorphic sections. Define
\begin{equation}
\label{dervs}
F_\alpha\eql {\partial F\over\partial X^\alpha}\,,\qquad F_{\alpha\beta}\eql {\partial^2 F\over\partial X^\alpha\partial X^\beta}\,.
\end{equation}
We can try to determine the prepotential by solving  a number of consistency condition that must be satisfied:

\begin{itemize}
\item [(i)] The holomorphic sections $X^\alpha$ and $F_\alpha$ are related by
\begin{equation}
\label{relxf}
F_\alpha\eql \cN_{\alpha\beta} X^\beta\,.
\end{equation}
From \eqref{holsect} and \eqref{Nmatr} we get
\begin{equation}
\label{fpreps}
F_0\eql -{i\over 2\sqrt 2}c_Xc_N(1-\zet^3)\,,\qquad F_1\eql {i\over 2}\sqrt{3\over 2}\,c_Xc_N\,\zet (1-\zet)\,.
\end{equation}
Note that $F_\alpha$'s came out holomorphic, just as they should have, which is by no means obvious looking at the explicit form of $\cN_{\alpha\beta}$.

\item [(ii)] One can now solve for the prepotential by exploiting its homogeneity and its degree:
\begin{equation}
\label{prepho}
F  \eql {1\over 2} X^\alpha F_\alpha\,, \qquad F\eql (X^0)^2 \,{\cal F}\left({X^1/ X^0}\right)
\end{equation}
By comparing the right hand side of \eqref{prepho}  written as functions of $z$, we obtain
\begin{equation}\label{eqforcF}
{\cal F}\left({\sqrt 3\,\zet(1+\zet)\over 1+\zet^3}\right)\eql -{i\over 8}{(1-\zet^2)^3\over (1+\zet^3)^2}\,,
\end{equation}
whose solution is
\begin{equation}
\label{solcF}
{\cal F}(x)\eql \pm {i\over 24}c_N\left(\sqrt 3+3\,x\right)^{1/2}\left(\sqrt 3-x\right)^{3/2}\,.
\end{equation}
Substituting this back into \eqref{prepho} and choosing one of the signs, we get
\begin{equation}
\label{prepot}
F(X^0,X^1)\eql  c_N\,{i\over 2 \cdot3^{3/4}}\left({1\over 2} \,X^0+{\sqrt 3\over 2}\,X^1\right)^{1/2}\left({\sqrt 3\over 2}\, X^0-{1\over 2}\, X^1\right)^{3/2}\,.
\end{equation}
We verify that the first relation in \eqref{dervs} is indeed satisfied, which confirms the consistency of the solution. 

The prepotential \eqref{prepot} is a rotation by $\pi/6$ of the prepotential, $\sqrt{X^0(X^1)^3}$,  known to be associated with the special geometry of the coset $\rm SU(1,1)/U(1)$, see, for example, \cite{deWit:1995jd}. 

\item[(iii)] The relative normalization between the sections, $X^\alpha$, and the matrix, $\cN_{\alpha\beta}$, is determined by
\begin{equation}
\label{norxandN}
{\rm Im}\,\cN_{\alpha\beta}\, X^\alpha\overline X{}^\beta\eql  -{1\over 2}e^{-K}\,,
\end{equation}
which gives
\begin{equation}
\label{norcon}
c_Nc_X^2|f_0|^2\eql c_Nc_L^2\eql 1\,.
\end{equation}
\item [(iv)] The normalization \eqref{norxandN} is related to the following expression for the K\"ahler potential in special geometry,
\begin{equation}
\label{kahler}
K\eql -\log i\left(\overline X{}^\alpha F_\alpha-X^\alpha\overline F_\alpha\right)\,,
\end{equation}
which is satisfied indeed.

\item [(v)] As a further consistency check, one can verify that the matrix, $\cN_{\alpha\beta}$, can be expressed in terms of the prepotential, $F$, and the holomorphic sections as follows:
\begin{equation}
\label{identN}
\cN_{\alpha\beta}\eql \overline F_{\alpha\beta} + 2 i {\left({\rm Im}\,F_{\alpha\gamma}\right)\left({\rm Im}\,F_{\beta\delta}\right)X^\gamma X^\delta\over \left({\rm Im}\,F_{\gamma\delta}\right)X^\gamma X^\delta}\,.
\end{equation}

\item [(vi)] Evaluating the vielbeins in  $f^\alpha_z$ in  \eqref{fvielb} we get
\begin{equation}
\label{vielex}
\begin{split}
f_z^0 & \eql  3\sqrt 2 \,c_L \,{\bzet+\zet^2\over (1-|\zet|^2)^{5/2}}\,,\\[6pt]
f_z^1 & \eql \sqrt 6\,c_L\,{1+2\zet+2\zet\bzet+\zet^2\bzet\over (1-|\zet|^2)^{5/2}}\,.
\end{split}
\end{equation}
The ratio of the two vielbeins above  is consistent with the ratio of variations in \eqref{varchi}. The difference in   powers in the denominators arises from different normalization of the spin-$1\over 2$ fields in  \cite{de Wit:1982ig}
and \cite{Andrianopoli:1996vr}, which is easily seen from the action. The spin-$1\over 2$ action in \cite{Andrianopoli:1996vr},
\begin{equation}
\label{spinhalffre}
e^{-1}\cL\eql -{1\over 2}i g_{\zet\bzet}\left(\bar\lambda^{\zet A} \gamma^\mu\nabla_\mu\lambda^{\bzet}{}_A+{\rm h.c.}\right)\,,
\end{equation}
contains explicit scalar metric, which is absent in the truncation of the corresponding term in \cite{de Wit:1982ig}.

\item[(vii)] In the calculation of the potential it is convenient to define a matrix
\begin{equation}
\label{Umatdef}
U^{\alpha\beta}\eql g^{\zet\bzet}f_z^\alpha f_{\bzet}^\beta\,,
\end{equation}
which is also equal to
\begin{equation}
\label{Umatother}
U^{\alpha\beta}\eql -{1\over 2} \,({\rm Im}\,\cN)^{-1\,\alpha\beta}-\overline L{}^\alpha L{}^\beta\,.
\end{equation}
It is gratifying to check that \eqref{Umatdef} and \eqref{Umatother} agree provided \eqref{norcon} holds.
\end{itemize}

\subsection{The quaternionic K\"ahler manifold}

The calculation of the quaternionic prepotentials, $\cP_\alpha^x$, is straightforward if one uses the fact that the coset $\rm SU(2,1)/SU(2)\times U(1)$ is a symmetric coset space. One then has the decomposition of the left-invariant form:
\begin{equation*}
g^{-1}dg\eql \omega^x h_x+\omega^0h_0+e^i t_i\,,
\end{equation*}
where $h_x$ are the generators of $\rm SU(2)$, $h_0$ is the $\rm U(1)$ generator and $t_i$ are the non-compact coset generators.  One can then read off the frames $e^i$ and the corresponding spin connection
\begin{equation}
\label{spincon}
\omega^i{}_j\eql \omega^xc_x{}^i{}_j+\omega^0 c_{0}{}^i{}_j\,,
\end{equation}
where $c_x{}^i{}_j$ and $c_{0}{}^i{}_j$ are the structure constants,
\begin{equation}
\label{strcons}
[h_x,t_i]\eql c_x{}^j{}_i t_j\,,\qquad [h_0,t_i]\eql c_0{}^j{}_i t_j\,.
\end{equation}
The coset generators, $t_i$, are the same as in \eqref{fundgen}. The invariant metric, $
 ds^2_{\rm QK}\eql e^i\otimes e^i\,, $ then
has the same normalization as the one in the kinetic scalar action \eqref{kinetic}.
The compact generators are rescaled such that they satisfy the standard commutators,
\begin{equation}
\label{comms}
[h_x,h_y]\eql   \epsilon^{xyz} h_z\,.
\end{equation}

Explicit expressions for the frames and  connection forms obtained using  coset representatives in \eqref{fundgen} are quite complicated when written in terms of the projective coordinates $\zeta_1$ and $\zeta_2$,\footnote{See,  \cite{Strominger:1997eb},  \cite{Behrndt:2000ph} and \cite{Gutperle:2001vw} for a different choice of coordinates and/or frames in which quantities associated with the quaternionic geometry are much simpler.} but are the natural ones given that the same coset representatives are used in the truncation.

The quaternionic K\"ahler forms, $\Omega^x$, are given by the curvatures of the $\rm SU(2)$ factor of the spin connection, 
\begin{equation}
\label{quatcurv}
 d\omega^x+{1\over 2}\epsilon^{xyz}\omega^y\wedge\omega^z\eql -\Omega^x\,.
\end{equation}
Using the Maurer-Cartan equations for $\rm SU(2,1)$ we find, 
\begin{equation}\label{theomforms}
\begin{split}
\Omega^x & \eql 2\left(e^1\wedge e^4+e^2\wedge e^3\right)   \,,\\[6pt] 
\Omega^y & \eql -2\left(e^1\wedge e^2+e^3\wedge e^4\right)\,,\\[6pt]
\Omega^z & \eql 2\left(e^1\wedge e^3-e^2\wedge e^4\right)\,.
\end{split}
\end{equation}
One verifies that tensors $J^x{}^i{}_j=\Omega^x_{ij}$, where $\Omega^x=\Omega^x_{ij}e^i\wedge e^j$, satisfy the algebra of quaternions,
\begin{equation}
\label{quatrel}
J^xJ^y\eql -\delta^{xy}+\epsilon^{xyz}J^z\,.
\end{equation}

Given a vector field $K$, its quaternionic moment map is a triple of functions, $\cP^x_K$,
satisfying
\begin{equation}
\label{defprep}
-i_K\,\Omega^x\eql d\,\cP^x_K+\epsilon^{xyz}\omega^y \,\cP^z_K\,.
\end{equation}
For a vector field, $K$, one can solve this equation algebraically using the  integrability condition. In this way we obtain the moment maps for the two Killing vector fields:
\begin{equation}
\label{momentsone}
\cP_{i\zeta_1\partial_{\zeta_1}+i\zeta_2\partial_{\zeta_2}+{\rm c.c.}} \eql
-{\left(1+|\zeta_{12}|^2\right)^2\over \left(1-|\zeta_{12}|^2\right)^2}
\left(
\begin{matrix}
  \zeta_1\bar{\zeta_2}+\zeta_2\bar{\zeta_1} \\[6pt] 
  i( \zeta_1\bar{\zeta_2}-\zeta_2\bar{\zeta_1}) \\[6pt]
   2\left( \zeta_{12} ^2+ \bar{\zeta}_{12} ^2\right) \left(1+|\zeta_{12}|^2\right)^{-2}
\end{matrix}
\right)\,,
\end{equation}
and
\begin{equation}
\label{momentstwo}
\cP_{i\zeta_1\partial_{\zeta_1}-i\zeta_2\partial_{\zeta_2}+{\rm c.c.}} \eql
-{\left(1+|\zeta_{12}|^2\right)^2\over \left(1-|\zeta_{12}|^2\right)^2}
\left(
\begin{matrix}
{1\over 2}( \zeta_1\bar{\zeta_2}+\zeta_2\bar{\zeta_1} )\left( \zeta_{12} ^2+ \bar{\zeta}_{12} ^2\right)\\[6pt]
 {i\over 2}( \zeta_1\bar{\zeta_2}-\zeta_{2}\bar{\zeta_1}) \left( \zeta_{12} ^2+ \bar{\zeta}_{12} ^2\right)\\[6pt]
  \left(2+ \zeta_{12} ^4+ \bar{\zeta}_{12} ^4\right) \left(1+|\zeta_{12}|^2\right)^{-2}\end{matrix}
\right)\,,
\end{equation}
where $\zeta_{12}$ is defined in \eqref{zetavar}.
This completes the list of ingredients needed for comparing our scalar potential with that of a $\cN=2$ theory.

\subsection{The scalar potential}

The scalar potential in gauged $\cN=2$ supergravity coupled to vector multiplets and hypermultiplets is given in\footnote{The equation number is different in the version on the {\tt arXiv}.}  (4.17) of   \cite{Andrianopoli:1996vr} in terms of geometric quantities associated with the special K\"ahler and quaternionic K\"ahler scalar manifolds. Specifying it to the field content here we have
\begin{equation}
\label{frepot}
V(\zet,\bzet;\zeta,\bar\zeta)\eql g^2\left(4\,g_{ij}K^i_\alpha K^j_\beta \overline L{}^\alpha L{}^\beta+g^{\zet\bzet}
f_\zet^\alpha f_{\bzet}^\beta \cP_\alpha^x\cP_\beta^x-3\overline L{}^\alpha L{}^\beta \cP^x_\alpha\cP^x_\beta\right)\,,
\end{equation}
where $K_\alpha$ are the Killing vectors of the gauged isometries in \eqref{killvect}, $P_\alpha$ are the corresponding moment maps and $g_{ij}$ is the metric on the quaternionic manifold. 
 
Note that, at least superficially, the structure of \eqref{frepot} is  quite similar to that of \eqref{poten}. Indeed, the ``$-3$" suggests that the last term should be the analogue of $-3|\cW|^2$. Then   \eqref{fvielb} implies  that $f_\zet ^\alpha\sim\partial_\zet L^\alpha$, while   \eqref{defprep} implies that $K_\alpha^i \sim \partial_i\cP^\alpha$, so that the first two terms are in a sense derivatives of the last one, just as in \eqref{poten}.   However, detailed calculations reveal the following:

\begin{itemize}
\item [(i)] Substituting \eqref{holsectL}, \eqref{momentsone} and \eqref{momentstwo} in \eqref{frepot} and using \eqref{superpot} and \eqref{superpotm}, we find
\begin{equation}
\label{fstcomp}
 \overline L{}^\alpha L{}^\beta\, \cP^x_\alpha\,\cP^x_\beta\eql 4 \,c_L^2\,\left(\left|\cW_+\right|^2+\left|\cW_-\right|^2\right)\,.
\end{equation}
This sets \begin{equation}\label{}
c_L=c_X\eql {1\over 2}\,,\qquad c_N\eql 4\,,
\end{equation}
thus fixing all constants (up to a sign) introduced above.

\item [(ii)] The naive expectation   that the first term (resp.\ second term) in \eqref{frepot} matches with the sum of the first terms in \eqref{poten} turns out incorrect. Indeed, one can evaluate those  terms setting  $\zeta_{12}=0$ to get
\begin{equation}
\label{frefst}
4\,g_{ij}K^i_\alpha K^j_\beta \overline L{}^\alpha L{}^\beta \Big|_{\zeta_{12}=\bar{\zeta}_{12}=0}\eql 0\,,
\end{equation}
but
\begin{equation}
\label{fstus}
{4 \over 3} (1 - |z|^2)^2\,\left[\,
\left|{\partial | \cW _+| \over \partial z}\right|^2+\left|{\partial | \cW _-| \over \partial z}\right|^2\,\right]_{\zeta_{12}=\bar{\zeta}_{12}=0}\eql 6\,{\left|\zet+\bzet^2\right|^2\over (1-|\zet|^2)^3}\,.
\end{equation}
However, 
\begin{equation}
\label{scndterm}
g^{\zet\bzet}
f_\zet^\alpha f_{\bzet}^\beta \cP_\alpha^x\cP_\beta^x \Big|_{\zeta_{12}=\bar{\zeta}_{12}=0}\eql  6\,{\left|\zet+\bzet^2\right|^2\over (1-|\zet|^2)^3}\,.
\end{equation}

\item [(iii)] At that point it is not that surprising to find that in general
\begin{equation}
\label{theeqs}
\begin{split}
&  {4 \over 3} (1 - |z|^2)^2\,\left[\,
\left|{\partial | \cW _+| \over \partial z}\right|^2+\left|{\partial | \cW _-| \over \partial z}\right|^2\,\right]+ (1 -
|\zeta_{12}|^2)^2  \left[\,\left|{\partial  | \cW_+ | \over \partial \zeta_{12}} \right|^2+\left|{\partial  | \cW _-| \over \partial \bar{\zeta}_{12}} \right|^2\,\right]\\[10 pt]
&\eql   4\,g_{ij}K^i_\alpha K^j_\beta \overline L{}^\alpha L{}^\beta +g^{\zet\bzet}
f_\zet^\alpha f_{\bzet}^\beta \cP_\alpha^x\cP_\beta^x  \,,
\end{split}
\end{equation}
which proves the equality of the two potentials.

\end{itemize}

We have thus completely recast the $\rm SU(3)$ truncation directly into the canonical language of special K\"ahler and quaternionic K\"ahler geometry.

\section{Branching rules}
\renewcommand{\theequation}{C.\arabic{equation}}
\setcounter{equation}{0} 
\label{appendixC}

In this appendix  we   collect some branching rules for  representations of $\rm SO(8)$ that  are used throughout the paper and in particular in the calculation of the scalar mass spectrum in Appendix~\ref{appendixA}. We use the same notation as in \cite{Warner:1983vz} and  write   $\mathbf{8}_\pm$  instead of the more common $\mathbf{8}_{s,c}$ (see, e.g.\ \cite{Slansky:1981yr}) to denote the two spinor representations of $\rm SO(8)$. The vector representation is denoted  as usual by $\mathbf{8}_v$. The reader should also note that our normalization of the $\rm U(1)$ charge is the same as in \cite{Nicolai:1985hs} and is by a factor of $-2$ smaller than the one in  Table 27 of \cite{Slansky:1981yr}.

All the branching rules that we need can be derived from the elementary tensor product
\begin{equation}
\mathbf{8}_{t} \otimes \mathbf{8}_{t} = \mathbf{35}_{t} \oplus \mathbf{28} \oplus \mathbf{1}\,,\qquad t=+\,, -\,, v\,.
\end{equation}
The 70 scalars of gauged supergravity are in the $\mathbf{35}_{-}$ and $\mathbf{35}_{+}$, in particular the pseudoscalars are in the $\mathbf{35}_{-}$. 

The embeddings  of $\rm SO(7)^\pm$, $\rm SU(4)^\pm$ and   $\rm SU(3)\times U(1)^{\pm}\subset SU(4)^\pm$ in $\rm SO(8)$ are defined by the following branching rules:
\begin{eqnarray}
\mathbf{8}_{\pm} & \underset{\rm SO(7)^{\pm}}{\longrightarrow}& \mathbf{7} \oplus \mathbf{1} \underset{\rm SU(4)^{\pm}}{\longrightarrow}  \left[ \mathbf{6} \oplus \mathbf{1} \right] \oplus \mathbf{1} \underset{\rm SU(3)\times U(1)^{\pm}}{\longrightarrow} \left[\left(\mathbf{\bar{3}}_{\frac{1}{3}} \oplus \mathbf{3}_{-\frac{1}{3}} \right) \oplus \mathbf{1}_{0}\right] \oplus \mathbf{1}_{0}~,\\\notag\\
\mathbf{8}_{\mp} &\underset{\rm SO(7)^{\pm}}{\longrightarrow}& \mathbf{8} \underset{\rm SU(4)^{\pm}}{\longrightarrow} \mathbf{4} \oplus \mathbf{\bar{4}} \underset{\rm SU(3)\times U(1)^{\pm}}{\longrightarrow} \left(\mathbf{3}_{\frac{1}{6}}\oplus \mathbf{1}_{-\frac{1}{2}} \right)\oplus \left( \mathbf{\bar{3}}_{-\frac{1}{6}} \oplus \mathbf{1}_{\frac{1}{2}}\right)~,\\\notag\\
\mathbf{8}_{v} &\underset{\rm SO(7)^{\pm}}{\longrightarrow}& \mathbf{8}  \underset{\rm SU(4)^{\pm}}{\longrightarrow} \mathbf{\bar{4}} \oplus \mathbf{4} \underset{\rm SU(3) \times U(1)^{\pm}}{\longrightarrow} \left( \mathbf{\bar{3}}_{-\frac{1}{6}} \oplus \mathbf{1}_{\frac{1}{2}}\right) \oplus \left(\mathbf{3}_{\frac{1}{6}}\oplus \mathbf{1}_{-\frac{1}{2}} \right)~.
\end{eqnarray}
Then the  branching rules for the three 35-dimensional representations of $\rm SO(8)$ are:
\begin{eqnarray}
\mathbf{35}_{\pm} & \underset{\rm SO(7)^{\pm}}{\longrightarrow} & \mathbf{27} \oplus \mathbf{7} \oplus \mathbf{1} \underset{\rm SU(4)^{\pm}}{\longrightarrow}  \left[\mathbf{20'} \oplus \mathbf{6} \oplus \mathbf{1} \right] \oplus \left[\mathbf{6} \oplus \mathbf{1}\right] \oplus \mathbf{1} \notag\\ &\underset{\rm SU(3)\times U(1)^{\pm}}{\longrightarrow}& \left[ \left(\mathbf{8}_{0} \oplus \mathbf{\bar{6}}_{\frac{2}{3}} \oplus \mathbf{6}_{-\frac{2}{3}} \right) \oplus \left( \mathbf{\bar{3}}_{\frac{1}{3}} \oplus \mathbf{3}_{-\frac{1}{3}} \right) \oplus \mathbf{1}_{0} \right] \oplus \left[ \left( \mathbf{\bar{3}}_{\frac{1}{3}} \oplus \mathbf{3}_{-\frac{1}{3}} \right) \oplus \mathbf{1}_{0} \right] \oplus \mathbf{1}_{0}\,,\quad \\\notag\\
\mathbf{35}_{\mp} &\underset{\rm SO(7)^{\pm}}{\longrightarrow}& \mathbf{35} \underset{\rm SU(4)^{\pm}}{\longrightarrow} \mathbf{15} \oplus \mathbf{\overline{10}} \oplus\mathbf{10} \notag\\ &\underset{\rm SU(3)\times U(1)^{\pm}}{\longrightarrow}& \left(\mathbf{8}_{0} \oplus \mathbf{\bar{3}}_{-\frac{2}{3}} \oplus\mathbf{3}_{\frac{2}{3}}\oplus \mathbf{1}_{0} \right) \oplus \left(\mathbf{\bar{6}}_{-\frac{1}{3}} \oplus \mathbf{\bar{3}}_{\frac{1}{3}} \oplus \mathbf{1}_{1}\right) \oplus \left(\mathbf{6}_{\frac{1}{3}} \oplus \mathbf{3}_{-\frac{1}{3}} \oplus \mathbf{1}_{-1}\right)\,,\\\notag\\
\mathbf{35}_{v} &\underset{\rm SO(7)^{\pm}}{\longrightarrow}& \mathbf{35}  \underset{\rm SU(4)^{\pm}}{\longrightarrow} \mathbf{15} \oplus \mathbf{\overline{10}} \oplus\mathbf{10} \notag\\ &\underset{\rm SU(3)\times U(1)^{\pm}}{\longrightarrow}& \left(\mathbf{8}_{0} \oplus \mathbf{\bar{3}}_{-\frac{2}{3}} \oplus\mathbf{3}_{\frac{2}{3}}\oplus \mathbf{1}_{0} \right) \oplus \left(\mathbf{\bar{6}}_{-\frac{1}{3}} \oplus \mathbf{\bar{3}}_{\frac{1}{3}} \oplus \mathbf{1}_{1}\right) \oplus \left(\mathbf{6}_{\frac{1}{3}} \oplus \mathbf{3}_{-\frac{1}{3}} \oplus \mathbf{1}_{-1}\right)\,.
\end{eqnarray}

On can also obtain the branchings with respect to $\rm SU(3)\times U(1)$ by going through $\rm G_2$. In that case the branching rule  for the three 8-dimensional representations of $\rm SO(8)$ are the same
\begin{equation}
\mathbf{8}_{\pm,v} \underset{\rm G_2}{\longrightarrow} \mathbf{7} \oplus \mathbf{1}  \underset{\rm SU(3)\times U(1)}{\longrightarrow}\left(\mathbf{\bar{3}}_{\frac{1}{3}} \oplus \mathbf{3}_{-\frac{1}{3}} \oplus \mathbf{1}_{0}\right) \oplus \mathbf{1}_{0}~,
\end{equation}
and we obtain 
\begin{equation}
\begin{split}
\mathbf{35}_{\pm,v} & \underset{\rm G_2}{\longrightarrow} \mathbf{27} \oplus \mathbf{7} \oplus \mathbf{1} \\
&\hspace{- 12 pt} \underset{\rm SU(3)\times U(1)}{\longrightarrow}  \left(\mathbf{8}_{0} \oplus \mathbf{\bar{6}}_{\frac{2}{3}} \oplus \mathbf{6}_{-\frac{2}{3}}  \oplus  \mathbf{\bar{3}}_{\frac{1}{3}} \oplus \mathbf{3}_{-\frac{1}{3}}  \oplus \mathbf{1}_{0} \right) \oplus  \left( \mathbf{\bar{3}}_{\frac{1}{3}} \oplus \mathbf{3}_{-\frac{1}{3}} \oplus \mathbf{1}_{0} \right) \oplus \mathbf{1}_{0}~,
\end{split}
\end{equation}
which agrees with the result above.

\end{appendices}



\end{document}